\documentclass{nature}

\usepackage{graphicx}
\makeatletter
\let\saved@includegraphics\includegraphics
\AtBeginDocument{\let\includegraphics\saved@includegraphics}
\renewenvironment*{figure}{\@float{figure}}{\end@float}
\makeatother

\newcommand{\um}{$\mu$m}
\newcommand{\arcsec}{''}

\title{The Magnetic Field Across the Molecular Warped Disk of Centaurus A}

\author{Enrique Lopez-Rodriguez$^{1,2}$}

\begin{document}

\maketitle

\begin{affiliations}
	\item{Kavli Institute for Particle Astrophysics \& Cosmology (KIPAC), Stanford University, Stanford, CA 94305, USA}
	\item SOFIA Science Center, NASA Ames Research Center, Moffett Field, CA 94035, USA
\end{affiliations}

\begin{abstract}
Magnetic fields are amplified as a consequence of galaxy formation and turbulence-driven dynamos. Galaxy mergers can potentially amplify the magnetic fields from their progenitors, making the magnetic fields dynamically important. However, the effect of mergers on magnetic fields is still poorly understood. We use thermal polarized emission observations to trace the magnetic fields in the molecular disk of the nearest radio active galaxy, Centaurus A, which is thought to be the remnant of a merger. Here, we detect that the magnetic field orientations in the plane of the sky are tightly following the $\sim3.0$ kpc-scale molecular warped disk. Our simple regular large-scale axisymmetric spiral magnetic field model can explain, to some extent, the averaged magnetic field orientations across the disk projected on the sky. Our observations also suggest the presence of small-scale turbulent fields, whose relative strength increases with velocity dispersion and column density. These results have strong implications for understanding the generation and role of magnetic fields in the formation of galaxies across cosmic time.
\end{abstract}

\section{Introduction} \label{sec:int}

Nearby galaxies are known to have regular large-scale magnetic fields (B-fields) with a spiral-like pattern at kpc-scales \cite{beck2019} but they also appear to have an important small-scale (or turbulent, or random) component \cite{BW2013,beck2019}. These regular large-scale B-fields are thought to be generated by a mean-field galactic dynamo, which relies on differential rotation of the galactic disk to amplify and order a `seed' B-field. This B-field is driven by supernova explosions at scales of $l \sim 50-100$ pc \cite{RSS1988,BS2005,Haverkorn2008}. The turbulent or random B-fields are thought to be generated by small-scale dynamo, which rely on turbulent gas motions at scales smaller than the energy-carrying eddies \cite{BS2005}. The correlation length of the turbulent or random B-fields is compared to or smaller than the turbulent scale, $l$. Once the B-fields are amplified and in close equipartition with thermal and turbulent forces, magnetic fields can influence galaxy evolution \cite{BS2005,PS2013,PMS2014,marinacci2018,su2018,ntormousi2018}. Galaxies at redshifts up to z $\sim2$, which are thought to be the progenitors of present-day galaxies, have been observed to host magnetic fields \cite{bernet2008,mao2017}. Magnetohydrodynamical simulations \cite{RT2016} suggest that during the violent feedback-dominated early phase in the galaxy formation history, weak seed B-fields can be first amplified by a small-scale dynamo. In the subsequent quiescent galaxy evolution phase, the turbulent or random B-fields can weaken or be maintained via large-scale dynamo action. These observations and simulations invite the question of the origin and evolution of the B-fields in galaxy evolution. 

Centaurus A is thought to be the remnant of a merger about $1.6-3.2 \times 10^{8}$ yr ago \cite{graham1979, struve2010}. The chaotic dust lane and shells within it form a warped disk of $\sim12$ kpc along the east-west direction with a median position angle of $122\pm4^{\circ}$ \cite{BM1954,graham1979} and inclination of $\sim90\pm30^{\circ}$ \cite{Q2010}. The observed parallelogram-shape structure in the mid-infrared (mid-IR) is caused by folds in a thin, dusty warped disk rich in molecular gas, which has the most active star formation present in the galaxy \cite{Q1992,mirabel1999,leeuw2002,Q2006a}. The warped disk has a rapidly rotating gas of $\sim 3$ kpc in radius based on the velocity field of several tracers \cite{vanGorkom1990}. A warped disk model was adequate to describe the mid-IR morphology and kinematics of the galaxy disk \cite{NBT1992,Q1992,QGF1993,S1996,Q2006a}. This model suggests that tidal forces during the merger modified the original gas motions of the spiral galaxy, forming rings around the central elliptical galaxy. In this scenario, the merger may enhance and amplify the B-fields by small-scale dynamo action when the B-fields are coupled to the gas component of the galaxy.

Pioneering work \cite{EH1964} using optical polarimetric observations with $24-69$-inch telescopes detected the polarization signature of dichroic absorption in the dust lane of Centaurus A. These authors concluded that the most likely reason for this detection is that the B-fields have been confined in the general orientation of motion of the gas in the galaxy disk. Although major efforts have been undertaken using optical and IR polarimetric studies, only small regions around the central active nucleus and  across the dust lane have been measured in the IR \cite{berry1985,bailey1986,hough1987,schreier1996,packham1996,scarrott1996,capetti2000,jones2000}. In general, the position angle (PA) of polarization is measured to be in the range of $110-117^{\circ}$, which is parallel to the dust lane and shows small angle fluctuations of $\sim9^{\circ}$ \cite{jones2000}. The degree of polarization (P) decreases from $\sim6$\% in the optical to $\sim2$\% in the IR, which is entirely consistent with polarization arising from dichroic absorption. These results indicate that the galaxy disk may have an ordered B-field, although the B-fields of the whole galaxy disk have not been traced yet using IR polarimetric techniques. Thus, a whole picture of the B-fields and how they relate to the gas dynamics in the molecular warped disk of Centaurus A is still missing.


\section{The Data} \label{sec:obs}

We observe Centaurus A using the High-resolution Airborne Wideband Camera-plus (HAWC+) on the $2.7$-m Stratospheric Observatory For Infrared Astronomy (SOFIA) telescope, with a beam size (full width at half maximum) of $7.80\arcsec$ at $89$ \um. We performed observations using the on-the-fly-map polarimetric mode (see Methods and Extended Data Figure 1).  Figure 1 shows the polarization map of Centaurus A at $89$ \um~observed with HAWC+ overlaid on a composite image, as well as with the total and polarized intensities at $89$ \um. The total flux image is consistent with the $70-160$ \um~\textit{Herschel} observations \cite{parkin2012}. The near-constant PA of polarization from optical to far-IR wavelengths indicates a single dominant polarization mechanism. We estimate that our 89 \um~polarization measurements arise from dichroic emission of magnetically aligned dust grains in the molecular disk (see Methods). 

The most prominent polarization signature of Centaurus A is the measured B-fields with orientations tightly following the $\sim3.0$ kpc-scale molecular warped disk. The measurements of the B-field orientations and warped disk morphology are those projected on the plane of the sky. We measure that the B-field orientations have a dispersion of $8.6-15.5^{\circ}$ across the observed $\sim181$\arcsec ~($\sim3.0$ kpc) molecular warped disk (see Methods and Extended Data Figure 2 and Extended Data Figure 3). However, the B-fields show a tightly orientation following the warped disk without systematic dispersion. The polarized flux morphology is spatially coincident with the low-surface brightness regions at the top and bottom of the edges of the parallelogram structure observed at $8-15$ \um~with \textit{Spitzer} \cite{Q2006a} and $70-160$ \um~with \textit{Herschel} \cite{parkin2012}. These regions are the closest to our line-of-sight (LOS) with low column density in the molecular disk \cite{parkin2012}, where the dust is optically thin at far-IR wavelengths.

We report the measurement of a polarized radio-loud active nuclei of $1.5 \pm 0.2$\% and PA of $151 \pm 4^{\circ}$ (B-field) within a 8\arcsec~(128 pc) diameter at 89 \um~(See Extended Data Figure 4). We estimated that the polarization arises from magnetically aligned dust grains, where the B-field orientation is found to be almost perpendicular to the radio jet axis, PA $\sim51^{\circ}$ \cite{israel1998}.

There are striking morphological similarities with the B-fields in the disks of highly inclined galaxies observed using radio polarimetric techniques \cite{Krause2020}. However, radio polarimetric observations of the host galaxy of Centaurus A have not been performed. We note that Faraday rotation is not a factor at far-IR wavelengths. The total gas column density is traced more effectively by our far-IR observations than by the relativistic electrons producing the synchrotron emission at radio. Near-IR polarization is subject to scattering effects from the disk, and dichroic absorption is only sensitive to the outer layers of the dust lane. Our far-IR observations trace deeper regions of the molecular disk than those at near-IR. In addition, far-IR emissive polarization observations reveal the orientation of the ordered B-field, but not its direction. Hence we cannot distinguish between regular large-scale fields and anisotropic random fields with frequent reversals.


\section{The large-scale regular magnetic field}\label{sec:largeBfield}

Spiral arms have been reported in the central $\sim3$ kpc of Centaurus A using $^{12}CO(2-1)$ observations \cite{E2012}, and no bar has been found using the HI 21 cm line \cite{struve2010}.  We produce a three-dimensional model of the regular large-scale B-field morphology using an axisymmetric spiral B-field configuration \cite{RG2010,BHB2010},  which is a mode of a galactic dynamo with a symmetric spiral pattern in the galactic midplane and a helical component (see Methods for a full mathematical description of the B-field model). This model is used to investigate trends from regular large-scale B-fields and not intended to truly represent the B-fields of a warped disk or to account for turbulent small-scale B-fields.

Our B-field morphological model (Figure 2) obtains an edge-on ($i=89.5^{+0.7}_{-0.8}$$^{\circ}$) galaxy with a tilt of $\theta=119.3^{+0.5}_{-0.4}$$^{\circ}$ east from north in the counter clockwise direction. The axisymmetric spiral B-field on the plane of the galaxy has a pitch angle of $\Psi_{0} = -54.9^{+0.5}_{-0.5}$$^{\circ}$, where the helical component has a radial pitch angle of $\chi_{0} = 74.8^{+10.5}_{-16.5}$$^{\circ}$ and vertical scale of $z_{0} = 1.7^{+0.3}_{-0.4}$ kpc. $\chi_{0}$ is defined as the pitch angle at a radius $z_{0}$ along the vertical axis of the galaxy (see Methods and Extended Data Figure 5 and Extended Data Figure 6 for details of the fitting routine). Our model results on a spiral B-field structure with a large pitch angle in the plane of the galaxy, potentially due to the tidal effects of the galaxy interaction. However, the molecular disk of Centaurus A is likely to be warped, where the inclinations, $i$, and tilted, $\theta$, angles have been measured \cite{Q2010} to be in the range of $[60,120]^{\circ}$, and $[92,169]^{\circ}$, respectively from 2 pc to 6500 pc. Specifically, the mean inclination and tilt angles are estimated to be $83\pm6^{\circ}$ and $114\pm14^{\circ}$ in the range of $[0.5-3]$ kpc \cite{Q2010}, respectively, in agreement with our inferred results. The complex dynamics along the disk may change the pitch angle, $\Psi_{0}$, as a function of the radius from the core.

Both our model and observations agree within $10^{\circ}$ along the mid-plane of the dust lane (Figure 3). We estimate the median difference between the PA of the B-fields of HAWC+, $PA_{H}$, and our model, $PA_{M}$ to be $\Delta PA = <PA_{H} - PA_{M}> = 3.6\pm22.7^{\circ}$. We find that the angular dispersions from our model and those from magnetically aligned dust grains at scales of $124.8$ pc resolution are greater than can be accounted for by errors from our observations ($\sigma_{PA} \le 9.6^{\circ}$) within the molecular disk (Figure 3). Thus, another B-field component (i.e. small-scale B-fields) is required to explain the angular dispersion between the regular axisymmetric B-field model and the measured B-field. 

Our model also shows a vertical and twisted pattern in the central region of Centaurus A at a PA $\sim30^{\circ}$, which is in close agreement with the radio jet at a PA $\sim51^{\circ}$ (East of North) at the core of the radio loud active nucleus. The central $\sim100$ pc of Centaurus A shows very complex structures \cite{E2009} that are expected to generate some level of misalignment between the radio jet axis and the kpc structures. It is also worth noting the change of the observed B-field orientation at $(X,Y) = (-0.6,+0.2)$ kpc in Figure 3, which may be explained by the change from the mid-plane spiral  pattern to the helical pattern. However, on the other edge at $(X,Y) = (0.0,+0.3)$ kpc, the helical pattern seems to have a lower effect. We point out that we have excluded the central $0.8 \times 0.8$ kpc$^{2}$ in our modeling because the polarization is affected by energetic processes associated with the AGN. The exclusion zone is determined by the low polarized regions from our observations and the physical inner bubble \cite{Q2006b}, which indicate that different mechanisms of polarization are taking place in the central $0.8 \times 0.8$ kpc$^{2}$. Thus, the aforementioned differences between our model and observations may be due to different physical mechanisms in play within the central 800 pc. The largest differences between our observations and model appear in the NW and SE edges of the warped disk. The observed B-field tends to be parallel to the dust lane at scales $>3$ kpc, and another B-field configuration may be needed at these scales.


\section{The thermal polarization and the multi-phase interstellar medium}\label{sec:smallBfield}

The B-field of the interstellar medium (ISM) in galaxies is turbulent, and its random fluctuations are not necessarily isotropic (i.e. anisotropic random B-fields). Note the different nomenclature in the literature: anisotropic random fields, tangled fields, striated, ordered random \cite{jaffe2019}, and our choice to follow the nomenclature by \cite{beck2019}. The B-fields in the ISM are typically described using a combination of ordered and random components, where the relationship between the fractional polarization and intensity (I) provides a proxy to characterize the effect of the turbulent component. In general, the fractional polarization in the ISM has been found to decrease with increasing column density \cite{jones1989,H1999}, which can be attributed to 1) variations in the alignment efficiency of dust grains with column density ($N_{H+H_{2}}$), 2) tangled B-fields along the LOS, and/or 3) turbulent fields. Note that the anisotropic random fields also contribute to the observed polarized dust emission.  As dynamos convert kinetic energy into magnetic energy, we use the measurements of the velocity dispersion as a proxy of the turbulence in the gas, where depolarization effects are expected if turbulence increases with increasing column density.

The warped disk of Centaurus A is rich in molecular gas, where the molecular gas emission generally originates in high-density regions of the ISM close to the spiral arms. We use the $^{12}CO(1-0)$ emission line to characterize the dynamics in the molecular disk of Centaurus A. The measurements of the velocity dispersion of the $^{12}$CO(2-1) emission line ($\sigma_{v, ^{12}CO(1-0)}$, see Extended Data Figure 7) observed by the Atacama Large Millimeter/submillimeter Array (ALMA) are used as a proxy of the turbulence in the molecular gas. We use the polarized intensity vs. total intensity ($PI-I$) plot to identify several physical regions in the galaxy disk of Centaurus A (Figure 4 and Extended Data Figure 8). We find three distinct regions: outer disk, molecular disk, and low polarized regions (see Methods for specific criteria). To quantify the effect of turbulent B-fields in these regions, we use the correlations between $P$, $I$, $PI$, temperature ($T$), $N_{H+H_{2}}$, and $\sigma_{v, ^{12}CO(1-0)}$ (Figure 5, and Methods for details of how these maps have been computed). We show the median values of these parameters for each region in Table \ref{tab:table1}.

For the whole galaxy, we find a relation between $P$, $T$, $N_{H+H_{2}}$, and $\sigma_{v, ^{12}CO(1-0)}$. We measure that $P$ decreases with $T$, and $N_{H+H_{2}}$, while the $I$ increases with $T$, and $N_{H+H_{2}}$. These trends imply an increase of turbulent fields and/or tangle B-fields as $N_{H+H_{2}}$ increases. This result is supported by the increase of $I$ and the decrease of $P$ with increasing $\sigma_{v, ^{12}CO(1-0)}$. The $PI$ remains almost constant with $T$, and $\sigma_{v, ^{12}CO(1-0)}$, with a marginal increase with $N_{H+H_{2}}$.

For the three regions, we find that each region has unique physical conditions (see Methods and Figure 5, Extended Data Figure 9). The outer disk of the galaxy, with a size of $\sim6$ kpc in diameter, is characterized by low $N_{H+H_{2}}$, $T$, $I$, and  $\sigma_{v, ^{12}CO(1-0)}$. We find that $N_{H+H_{2}}$ and $\sigma_{v, ^{12}CO(1-0)}$ are correlated with high $P$. We conclude that the most plausible explanation is that an optically thin layer of diffuse ISM, with low velocity dispersion and ordered B-field, may be the main physical component of the measured B-field.

The molecular disk, of a size of $\sim3$ kpc in diameter, has a large $\sigma_{v,^{12}CO(1-0)} = 18.4\pm9.2$ km s$^{-1}$ (see Extended Data Figure 10) in comparison with the thermal velocity dispersions of $\sim8$ km s$^{-1}$ for molecular clouds in nearby galaxies \cite{CP2013}. The measured velocity dispersion may arise from multi-components along the LOS due to large velocity gradients in the molecular disk. This region may be affected by 1) averaging out many B-field orientations along the LOS across the molecular disk, and 2) high turbulence at the high density regions in the molecular disk. 

The low polarized regions with high $N_{H+H_{2}}$ and $I$ are found to have large $\sigma_{v,^{12}CO(1-0)} = 34\pm4$ km s$^{-1}$, which are spatially associated with the circumnuclear disk within the central $500$ pc and the active nucleus. This region is spatially coincident with the inner bubble \cite{Q2006b}, which is dominated by high energetic processes from the AGN. This region may be affected by 1) an increase of  velocity dispersion of the gas due to AGN activity, 2) a decrease of the dust grain alignment efficiency due to the large turbulence fields, and/or 3) competing mechanisms of polarization (i.e. B and/or K radiative torque alignment).  At the location of the active nucleus,  an ordered B-field in a dusty circumnuclear disk around the active nucleus may be the dominant physical structure producing the measured far-IR polarization.

\section{The origin of the magnetic field in Centaurus A}\label{Borigin}

Our B-field model at radius $>500$ pc reproduces the ordered B-fields parallel to the dust lane of the galaxy from optical absorptive polarization observations \cite{EH1964,berry1985,bailey1986,schreier1996,scarrott1996,capetti2000}. Our interpretation is that the outer layers of the dust lane may have a less turbulent B-field than at deeper regions of the galaxy, where the large velocity dispersion in the molecular gas may be enhancing the turbulent B-field in the molecular warped disk. Our model also reproduces, to some extent, the orientations of the observed B-fields along the central $\sim3$ kpc of the galaxy from our far-IR observations. We find that the morphology given by the thermal emission of the warped disk deviates from our regular large-scale B-field model. The inferred B-field orientations from our observations have higher spatial correspondence with the structure of the warped disk than with the regular large-scale B-field model. It is difficult to combine the regular large-scale B-field model with warped ring models \cite{NBT1992,Q1992,QGF1993,S1996,Q2006a} to 
explain the warped disk. Furthermore, other physical mechanisms, i.e. small-scale turbulent fields, may play an important role in the B-fields of the warped disk of Centaurus A. 

We show that our measured angular dispersions are larger than those purely arising from the observational uncertainties. In addition, we find that the polarized emission is less affected by the $T$, $N_{H}$, and $\sigma_{v,^{12}CO(1-0)}$ than the unpolarized emission. Our interpretation is that the isotropic turbulent field, traced by the unpolarized thermal emission, is more affected by these quantities than the ordered field, where the latter consists of the regular large-scale and anisotropic small-scale components. A possible explanation is that the small-scale turbulent field is relatively more significant at higher velocity dispersions of the molecular gas and column densities than the large-scale axisymmetric field. Our results can be interpreted as a decreasing ratio of the large-to-small B-fields. Thus, a substantial amount of the observed B-field at far-IR wavelengths may arise from anisotropic small-scale turbulent (ordered) fields that also contribute to the polarized emission or tangled fields at scales below or beyond the $124.8$ pc from our observations. 

The most likely scenario is that the observed B-fields have been generated by a dominant small-scale dynamo across the fast rotating and turbulent gas and dust molecular warped disk. For the outer disk, the turbulence may be driven by disk gravitational instabilities mixed with density wave or merge-driven streaming motions. In addition to the turbulent B-fields from the dense and cold ISM of spiral galaxies, galaxy interaction can enhance this turbulence, and produce disordered gas flows and shear. These gas flows can enhance and produce highly anisotropic turbulent fields and weak regular fields \cite{CB2004,D2011}, and regenerate the B-fields at a faster rate than those observed in spiral galaxies \cite{RS2016}. Differential rotation by the high rotational velocity field, $\sim280$ km s$^{-1}$ \cite{vanGorkom1990,Q1992}, in the molecular disk may also contribute to regenerate the B-fields at a fast rate  \cite{RSS1988}.

The B-field orientations within the central $\sim500$ pc diameter, excluding the active nuclei, from our $89$ \um~emissive polarization differs from the $1-2$ \um~absorptive polarization orientations \cite{packham1996,jones2000}. This difference can be explained by the far-IR emissive polarization tracing deep regions of the galaxy disk. In addition, this region is also affected by the inner bubble \cite{Q2006b}. Due to extinction effects, the near-IR absorptive polarization is only sensitive to the outer layers of the dust lane, which is not affected by the dynamics of the molecular warped disk or the inner $\sim500$ pc bubble. Interestingly, our B-field model within the central $500$ pc shows an ordered pattern that is more compatible with the near-IR absorptive observations than the far-IR emissive observations. 

At the location of the active nucleus, we measured a polarized point source with a B-field orientation almost perpendicular to the direction of the jet. We interpret that the B-field may arise from a circumnuclear dusty structure around the obscured active nuclei. These physical conditions are similar to those observed in the radio-loud Cygnus A galaxy \cite{LR2018b}.

We have demonstrated that far-IR polarization observations are a powerful tool to study the magnetic field morphology in the cold ISM of galaxies, especially when radio polarimetric observations are difficult due to contamination by radio jets or other radio sources. Due to the $180^{\circ}$ ambiguity from our observations, we cannot study the B-field direction, so the regular large-scale B-field direction must be analyzed with future Faraday rotation measurements obtained from radio polarimetric observations. These observations are required to distinguish between the compressed, amplified, highly anisotropic fields and unidirectional field generated by the dynamo process. In addition, further hydromagnetical simulations are required to refine our results to explain a regenerated field by the merger of an elliptical and spiral galaxy. This work serves as a strong reminder of the potential importance of B-fields, usually completely overlooked, in the formation and evolution of galaxies.



\begin{addendum}
 \item We sincerely thank Kandaswamy Subramanian, Kostas Tassis, Ric Davies, B-G Andersson, and Tom Osterloo for many useful discussions on theoretical approaches, hydromagnetic simulation, gas dynamics, and dust grain alignment theories. We are grateful to the four anonymous referees, whose comments greatly helped to clarify and improve the manuscript.  Based on observations made with the NASA/DLR Stratospheric Observatory for Infrared Astronomy (SOFIA) under the 07\_0032 Program. SOFIA is jointly operated by the Universities Space Research Association, Inc. (USRA), under NASA contract NAS2-97001, and the Deutsches SOFIA Institut (DSI) under DLR contract 50 OK 0901 to the University of Stuttgart. 
 \item[Author Contributions] E.L.-R.  is responsible for all aspects of this paper, i.e. lead the project, carried out observations, developed the analysis methods and data reductions, interpreted results, and wrote the text.
 \item[Competing Interests] The author declares that he have no competing financial interests.
 \item[Correspondence] Correspondence and requests for materials should be addressed to E.L.-R.~(email: enloro@gmail.com).
\end{addendum}

\begin{addendum}
\item[Figure 1:] \textbf{The measured B-field of Centaurus A using far-infrared polarimetric observations with HAWC+/SOFIA.} 
\textbf{a},  Composite image with overlaid normalized B-field orientation (black) and total intensity contours (blue) using  89 \um~SOFIA/HAWC+ observations. The FOV of the image is $600\arcsec \times 400\arcsec$ ($9.6 \times 6.4$ kpc$^{2}$).  The composite image has 870 \um~observations from LABOCA/APEX (orange), X-ray data from  \textit{Chandra} (blue), and V-band data from the Wide field Imager on the MPG/ESO 2.2-m telescope at La Silla, Chile showing the background stars and the dust lane.
\textbf{b}, 89 \um\ total flux (color scale) with overlaid polarization measurements (white) with $P/\sigma_{P} > 2.5$ and rotated by $90^{\circ}$ to show the measured B-field orientation within the central $250\arcsec \times 180\arcsec$ ($4.0 \times 2.9$ kpc$^{2}$) region. Contours are shown in steps of $2^{n}\sigma$, where $n = 2.0, 2.5, 3.0, \dots$ and $\sigma = 1.06$ mJy sqarcsec$^{-1}$.  
\textbf{c}, Polarized flux (color scale) with total intensity contours and B-field orientation as image on \textbf{b}. A legend polarization of $10$\% (black) and beam size of $7.8$\arcsec (red circle) are shown.
\item[Figure 2:] \textbf{Three-dimensional representation of the best inferred B-field morphological model of Centaurus A.}
\textbf{a}, The B-field morphology within $11\times11$ kpc$^{2}$ from a face-on view, 
\textbf{b}, at the inclination and tilt angle as inferred from our model, 
\textbf{c}, and within the central $8\times3$ kpc$^{2}$ are shown. The XYZ axes are shown for each panel. A fake colormap has been added to visually distinguish the B-field lines as a function of the distance to the core.
\item[Figure 3:] \textbf{B-field model vs. observations.} 
\textbf{a}, Best inferred B-field model (black lines) at the same angular resolution ($3.90$\arcsec) as the HAWC+ observations at 89~\um~(red circle). The measured B-field (blue lines) by HAWC+ is shown. Low polarized emission and those within the central $0.5$ kpc have been masked to only account for the magnetic field in the molecular disk.  \textbf{b}, Angular difference between the B-field orientations from HAWC+, PA$_{H}$, and model, PA$_{M}$. Each pixel is at the Nyqvist sampling ($3.90$\arcsec) and central polarization measurements have been masked as described in Section \ref{sec:largeBfield}. The red circle (bottom right) shows the beam size of the HAWC+ observations. 
\textbf{c},  Histogram of the angular difference in bins of $5^{\circ}$. The median (dashed vertical line) and $1\sigma$ (thin-dashed vertical lines) angular differences are shown.
\item[Figure 4:] \textbf{Physical regions of Centaurus A as a function of their polarization properties.} 
\textbf{a}, $PI-I$ plot vs. velocity dispersion of the $^{12}CO(2-1)$ emission line from ALMA observations (colorscale). The blue dotted vertical lines at $I = 1000$ and $2700$ MJy sr$^{-1}$ show the limits of the three physical regions found in this analysis.  The dashed diagonal black lines show the maximum expected polarization, $P\propto I^{0}$ = 15, 6.5, and 1.5\% for each of these physical regions, respectively. The $1-\sigma$ uncertainty of the polarized fluxes are shown.
\textbf{b}, The three physical regions, outer disk (black), molecular disk (blue), and low polarized regions (red), are shown. The measured B-field (white lines) is shown as in Figure 1.
\item[Figure 5:] \textbf{Polarization measurements as a function of the multi-phase ISM.} $P$, $I$, and $PI$ vs. $T$, $N_{H+H_{2}}$, and $\sigma_{v,^{12}CO(1-0)}$ for the outer disk (black circles), molecular disk (blue squares), and low polarized regions (red crosses) identified on Figure 4. For all figures, the $1-\sigma$ uncertainties are shown. A weighted power-law fit is shown for each region with the power-law indexes in Extended Data Figure 9.
\item[Table 1:] \textbf{Medians of the physical parameters of each region identified in Figure 4.}
\end{addendum}

\begin{methods}\label{sec:methods}

\subsection{Centaurus A.}  Another alternative name is NGC 5128. The galaxy disk divides the elliptical galaxy and obscures the nucleus and most of the structures at optical wavelengths within the central $\sim$500 pc \cite{schreier1996}.  Estimations of the velocity field of several tracers are HI \cite{vanGorkom1990} H$_{\alpha}$ \cite{BTA1987}, $^{12}$CO(1-0) \cite{eckart1990}, and $^{12}$CO(2-1) \cite{phillips1987}. The merger is estimated to be about $1.6-3.2 \times 10^{8}$ yr ago \cite{graham1979,thomson1992,NBT1992,Q1992,QGF1993, struve2010}.

\subsection{Distance.} We have taken a distance of $3.42\pm0.18$ (random) $\pm0.25$ (systematic) Mpc by \cite{ferrarese2007}. The distance is estimated using Cepheid variables by the \textit{Hubble Space Telescope}.  Other authors quote $3.63\pm0.07$ Mpc \cite{karachentsev2002} based on median distance to the galaxy group formed by 17 galaxies (Centaurus A included); and  $3.84\pm0.25$ Mpc \cite{rejkuba2003} using two independent methods: Mira period luminosity and the tip of the red-giant branch. In general, the range of distances is estimated to be $3.2-4.2$ Mpc based on the review by \cite{israel1998}. We estimate that any potential uncertainties ($\sim21.8$ pc) due to the estimation of the distance to Centaurus A are smaller than a single detector pixel scale ($\sim70$ pc) from our observations.

\subsection{Observations and Data Reduction.} Centaurus A was observed (PI: Lopez-Rodriguez, E., ID: 07\_0032) at $89$ \um~using the High-resolution Airborne Wideband Camera-plus (HAWC+) \cite{Vaillancourt2007,Dowell2010,Harper2018} on the $2.7$-m Stratospheric Observatory For Infrared Astronomy (SOFIA) telescope. Observations were performed on 20190717 during the SOFIA New Zealand deployment. HAWC+ polarimetric observations simultaneously measure two orthogonal components of linear polarization arranged in two arrays of $32 \times 40$ pixels each, with a pixel scale of  $4.02\arcsec$ pixel$^{-1}$, and beam size (full width at half maximum, FWHM) of $7.80\arcsec$ at $89$ \um. We performed observations using the on-the-fly-map (OTFMAP) polarimetric mode. This technique is an experimental observing mode performed during SOFIA Cycle 7 observations as part of engineering time to optimize the polarimetric observations of HAWC+.  We here focus on the scientific results of Centaurus A and describe the high-level steps of these observations.

We performed OTFMAP polarimetric observations in a sequence of four Lissajous scans, where each scan has a different halfwave plate (HWP) position angle (PA) in the following sequence: $5^{\circ}$, $50^{\circ}$, $27.5^{\circ}$, and $72.5^{\circ}$. This sequence is called `set' hereafter. In this new HAWC+ observing mode, the telescope is driven to follow a parametric curve with a nonrepeating period whose shape is characterized by the relative phases and frequency of the motion. An example of the OTFMAP for total intensity observations of NGC 1068 using HAWC+/SOFIA is shown by \cite{LR2018a}. Each scan is characterized by the scan amplitude, scan rate, scan phase, and scan duration.  A summary of the observations is shown in Extended Data Figure 1. The scan amplitude is defined by the length of the scan parallel (EL) and perpendicular (XEL) to the direction of the telescope elevation. We performed rectangular scans to cover the central molecular warped disk (the `parallelogram') of $\sim3.8$ kpc ($\sim 240\arcsec$) in diameter along the diagonal of the scan. 

We reduced the data using the Comprehensive Reduction Utility for SHARP II v.2.42-1 (\textsc{crush}\cite{kovacs2006,kovacs2008}) and the \textsc{HAWC\_DRP\_v2.3.2} pipeline developed by the data reduction pipeline group at the SOFIA Science Center. Each scan was reduced by \textsc{CRUSH}, which estimates and removes the correlated atmospheric and instrumental signals, solves for the relative detector gains, and determines the noise weighting of the time streams in an iterated pipeline scheme. Each reduced scan produces two images associated with each array. Both images are orthogonal components of linear polarization at a given HWP PA.  We estimated the Stokes IQU parameters using the double difference method in the same manner as the standard chop-nod observations carried by HAWC+ described in Section 3.2 by \cite{Harper2018}. The degree (P) and PA of polarization were corrected by instrumental polarization (IP) estimated using OTFMAP polarization observations of planets. We estimated an IP of Q/I $= -2.1$\% and U/I $= 0.8$\% at $89$ \um, with an estimated uncertainty of $\sim0.8\%$. The IP using OTFMAP observations are in agreement within their uncertainties with the estimated IP using chop-nod observations of Q/I $= -1.6$\% and U/I $= 0.8$\%. To ensure the correction of the PA of polarization of the instrument with respect to the sky, we took each set with a fixed line-of-sight (LOS) of the telescope. For each set, we rotated the Stokes QU from the instrument to the sky coordinates. The polarization fraction was debiased and corrected by polarization efficiency. The final Stokes IQU, P, PA, polarized intensity (PI), and their associated errors were calculated and re-sampled to one-quarter of the beam size, $1.95\arcsec$ at $89$ \um.

Several advantages and limitations are found with the OTFMAP polarization mode. The  advantages are the reduction of overheads and radiative offsets when compared with the chop-nod technique. The overheads are improved by a factor of two in comparison with the chop-nod technique. This improvement is due to the fact that the OTFMAP is constantly integrating with the source on the FOV while covering off-source regions to estimate the background levels. For the OTFMAP method, the telescope is always on-axis, without chopping the secondary mirror as it is in the chop-nod technique. Thus, the radiative offset is not present and the sensitivity of the observations was estimated to improve by a factor of 1.6. The limitation of this technique is the recovering of large-scale diffuse and faint emission from the astrophysical objects. This is a result of the finite size of the array, variable atmosphere conditions, variable detector temperature, and the applied filters in the reduction steps to recover extended emission. In general, the noise increases as a function of the length, $L$, of the extended emission as $\sim L^{2}$. Although we can adjust \textsc{crush} parameters to recover the extended emission in total intensity, the polarization is highly affected by the filter selection of the data reduction software. We applied several filters using \textsc{crush} to recover large-scale emission structures of Centaurus A without compromising the intrinsic polarization pattern of the astrophysical object. We performed chop-nod and OTFMAP observations of well-known objects, e.g. 30 Doradus and OMC-1, to test several filter options. We conclude that a combination of the extended filter with 12 iterations using \textsc{crush} can recover large-scale emission structures up to $200\arcsec$~from our observations of Centarus A at $89$ \um. Using \textit{Herschel} images at $70$ and $160$ \um~\cite{parkin2012}, we estimate that the fluxes at the regions where we are not able to recover large-scale emission from our observations are $\le 0.01$ Jy/sqarcsec at $89$ \um. For the total on-source time of $3200$s (Extended Data Figure 1) and assuming an expected degree of polarization of $\sim5$\% from our observations at $89$ \um, the signal-to-noise ratio (SNR) of the polarization is estimated to be $\le1.5$ for the unrecovered regions (Sensitivities can be estimated using the SOFIA Instrument Time Estimator (SITE) at https://dcs.arc.nasa.gov/proposalDevelopment/SITE/index.jsp). Although we are missing some of the large-scale structures, our observations would have not been sensitive enough to provide statistically significant polarization measurements of this region. As the molecular disk of Centaurus A has a size of  $\sim140\arcsec$ in diameter at $89$ \um, we are able to achieve our scientific goals using our observations. 

\subsection{Physical regions as a function of the B-field orientation and its dispersion.}  We computed histograms of the degree of polarization and B-field orientations for all polarization measurements with $P/\sigma_P \ge 3.0$. We identify three regions in the galaxy disk based on the overall orientation and dispersion of the magnetic field: 1) west side with a PA range of $[90,120]^{\circ}$, 2) east side with a PA range of $[120,175]^{\circ}$, and 3) low-polarized regions with large angle dispersion at PA $>175^{\circ}$. For each region, we compute the mean of the degree of polarization and B-field orientation and their dispersions (Extended Data Figure 2). Extended Data Figure 3 shows a color code polarization map of the three regions. 

\textit{The west and east regions of the warped disk:} We estimate a median PA of the B-field orientation of $105^{\circ}$ in the west side of the galactic disk, and $147^{\circ}$ in the east side. Previous optical to near-IR polarimetric studies have measured a PA of polarization of $110-117^{\circ}$ from dichroic absorption, which is roughly parallel to the dust lane \cite{berry1985,bailey1986,hough1987,schreier1996,packham1996,scarrott1996,capetti2000,jones2000}. These measurements were based on observations of several patches across the dust lane and/or the central $\sim1$ kpc.  

\textit{The central kpc:} The third region is mostly spatially coincident with low polarized areas within the central $\sim0.8$ kpc diameter around the AGN and at the edge of the northwest regions of the galaxy disk. This region has the largest PA dispersion of $28.9^{\circ}$. Extended Data Figure 4 shows a zoom-in of the central $50\arcsec \times 50\arcsec$ ($0.8 \times 0.8$ kpc$^{2}$), where the twist of the PA of polarization from the dust lane to the nucleus can be seen. The central $0.8$ kpc region lacks molecular gas and dust due to energetic processes near the AGN that have disturbed the inner regions of the warped disk \cite{Q2006a,Q2006b}. We conclude that the large PA dispersion, low polarization, and change of PA of polarization are due to the combination of the intrinsic polarization in the dust lane and the nucleus, both having a different PA of polarization and arising from different physical structures.

\subsection{Magnetic field model.} In this section, we describe the mathematical description of the magnetic field  model. The three components of the vectorial magnetic field in cylindrical coordinates ($r$, $\rho$, $z$) centered at the galactic center are described as

\begin{eqnarray}
B_{r}      & = &  B_{0} \sin \Psi_{0} \cos \chi_{z} \\
B_{\rho} & = &  B_{0} \cos \Psi_{0} \cos \chi_{z} \\
B_{z}     & = &  B_{0} \sin \chi_{z}
\end{eqnarray}
\noindent
where $B_{0}$ is the amplitude of the regular magnetic field strength, $\Psi_{0}$ is the pitch angle of the spiral pattern, and $\chi_{z}$ is  the pitch angle vertical field given by 

\begin{eqnarray}
\chi_{z} = \chi_{0} \tanh (\frac{z}{z_{0}})
\end{eqnarray}
\noindent
where $\chi_{z}$ is interpreted as a helical angle with a radial pitch angle $\chi_{0}$, purely longitudinal with a given vertical scale at  $z_{0}$ \cite{Laing1981}. 

The magnetic field is projected on the plane of the sky, which adds two free parameters, the inclination $i$, and tilt angle $\theta$. For a face-on view $i = 0^{\circ}$, and an edge-on view $i = 90^{\circ}$. The tilt angle, also described as the position angle of the major axis of the projected galaxy plane, has a reference $\theta = 0^{\circ}$ along the north-south direction and positively increases east from north. We use Euler rotations around the x-axis, $R_{x}[i]$, and the z-axis, $R_{z}[\theta]$, to compute the final magnetic field $B_{sky} = R_{x}[i]R_{z}[\theta]\textbf{B$_{o}$}$, where \textbf{B$_{o}$} = ($B_{x}$, $B_{y}$, $B_{z}$) in cartesian coordinates. 

\textit{Model constraints.} Thermal emissive polarization arising from magnetically aligned dust grains is not directly sensitive to the magnetic field strength but rather to variations in dust grain alignment, and gradients in temperature and column density. Although we can include a variation of the magnetic field strength as a function of the radius in our model, this information is negligible for interpreting the thermal emissive polarization. Radio polarimetric observations are sensitive to the magnetic field strength, but these observations have not been acquired for the warped disk of Centaurus A. Thus, we do not have any information about the radial dependency of the magnetic field strength in Centaurus A, where $B_{0}$ is an unknown variable with no constraints. If we still include a variable magnetic field strength as a function of the radius such as $B_{0} = r^{-\alpha}$ and assume an isotropic strength variation on the XYZ axes, given that the projected PA of polarization on the plane of the sky from our model is $PA_{B} = \arctan{(B_{y,sky}/B_{x,sky})}$, $PA_{B}$ does not depend on the value of $B_{0}$. A model using an isotropic variation of the magnetic field strength produces the same B-field orientation as our current model. If we consider the radial dependence of the B-field to be different for each  XYZ axes, then PA$_B$ changes as a function of the radius. This model provides a different B-field orientation when compared with our model. As we do not have information about the radial dependency of the magnetic field strength in Centaurus A,  and we are only interested in the orientation of the magnetic field, we assume a constant magnetic field strength, $B_{0} = const.$ 

The polarization in the central $0.8 \times 0.8$ kpc$^{2}$ is affected by energetic processes associated with the AGN. Therefore, we have excluded this region in this analysis. The exclusion zone is determined by the low polarized regions (Extended Data Figure 3) and the physical inner bubble found by \cite{Q2006b}, which indicate a different mechanism of polarization than the one taking place in the galactic disk.

\textit{Computation of the magnetic field model.} These definitions allow us to explore the parameter space of the five free parameters $\Psi_{0}$, $\chi_{0}$, $z_{0}$, $i$, and $\theta$. We compute synthetic magnetic field orientation maps, as well as images of the expected distribution of the magnetic field orientation in a full three-dimensional view of the magnetic field morphology of Centaurus A. We compute the magnetic field morphology within a box of $231\times176\times10$ pixels$^{3}$ with a physical pixel scale of $64$ pc equal to the Nyqvist sampling of the $89$ \um~HAWC+ observations, which corresponds to a physical volume of $14.78 \times 11.26 \times 0.64$ kpc$^{3}$. We perform a Markov Chain Monte Carlo (MCMC) approach using the dynamical evolution Metropolis sampling step in the \textsc{python} code \textsc{pymc3} \cite{pymc3}. The prior distributions are set to flat within the ranges shown in Extended Data Figure 5. We performed a blind exploration for $\Psi_{0}$ and $\chi_{0}$ in the $[-90,90]^{\circ}$ range and found that $\Psi$ tends to be negative and $\chi$ positive, so we constrained the prior distributions with the ranges shown in Extended Data Figure 5. We run 10 chains each with 5000 steps and an additional burning 2000 steps, which provides a total of 50000 steps for the full MCMC code useful for data analysis.  Final posterior distributions, median values, and $1\sigma$ uncertainties are shown in Extended Data Figure 5 and Extended Data Figure 6.

\textit{Model vs. Observations.} \cite{E2009,E2012} measured a tilt angle, $\theta$, of $\sim120^{\circ}$ for the $\sim1$ kpc central region of the $^{12}CO(2-1)$ emission, in agreement with our inferred $\theta = 119.3^{+0.5}_{-0.4}$$^{\circ}$. Using $^{12}CO(2-1)$ emission, \cite{E2012} found a spiral arm from $0.2$ kpc to $\sim1$ kpc, with a width of $0.5\pm0.2$ kpc and pitch angle of $-20^{\circ}$. These authors fitted by eye a logarithmic spiral to the deprojected $^{12}CO(2-1)$ emission using an inclination angle of $70^{\circ}$. Our model uses a three-dimensional axisymmetric spiral pattern with a helical component of the magnetic field, which inferred an inclination close to an edge-on view $i = 89.5^{+0.7}_{-0.8}$$^{\circ}$. The discrepancy of pitch angles may arise in the differences between model selection and projection effects. We tested our models with low inclinations to compare with the results from \cite{E2012}. In the range of $[70-80]^{\circ}$, this set of models provide reasonably good inferred magnetic field configurations only for the NW region of the molecular disk. For these models, we measured an angular offset $\Delta PA = PA_{H} - PA_{M} > 30^{\circ}$ for all polarization measurements in the SE region of the molecular disk. Thus, all these solutions are not considered in the final inferred magnetic field configuration presented above.

If we consider SNR$_{PA} = <PA_{H} - PA_{M}>/\sigma_{PA_{H}}$ as a measurement of the signal-to-noise between the residuals and the observational uncertainties, we estimate that SNR$_{PA}$ is in the range of $[2.4-10.4]\sigma$ level, where $2.4\sigma$ means that the residual is $2.4$ times larger than the observational PA uncertainty. We estimate that $\sim25.7$\% ($78$ out of $303$) of measurements are within $[2.4-3.0]\sigma$.

\textit{Three-dimensional B-field model.} As we compute a full three-dimensinal magnetic field model, we have produced a representation of the magnetic field lines for the best inferred model in Figure 2 using the 3D data visualization \textsc{python} package \textsc{Mayavi} (Mayavi can be found at https://docs.enthought.com/mayavi/mayavi/) \cite{mayavi}. 

\subsection{Alternative magnetic field models.} In this section, we describe the alternative B-field models. We compare our magnetic field model with other alternative configurations. Bisymmetric magnetic field configurations can also be used to describe the morphologies of magnetic field in galaxies. However, this morphology has only been argued for M81, which may probably be affected by Faraday depolarization \cite{beck2019}. We study this magnetic field configuration described as \cite{RG2010}

\begin{eqnarray}
B_{r}      & = &  B_{0} \cos{[\phi \pm \beta \ln(\frac{r}{r_{0}})]} \sin \Psi_{0} \cos \chi_{z} \\
B_{\rho} & = &  B_{0} \cos{[\phi \pm \beta \ln(\frac{r}{r_{0}})]} \cos \Psi_{0} \cos \chi_{z} \\
B_{z}     & = &  B_{0} \sin \chi_{z}
\end{eqnarray}
\noindent
where $\phi$ is the azimuthal angle, $\beta = 1/\Psi_{0}$ \cite{FT1980}, and $r_{0}$ is the radial scale of the pitch angle in the plane of the galaxy. For Milky Way studies, typically, $r_{0}$ is fixed at the Sun galactrocentric distance for a given measurement of the magnetic field strength. However, we do not have this information for Centaurus A and $r_{0}$ is an extra free parameter, which we assume to be in the range of $[0,10]$ kpc in our model. We performed the same fitting methodology as described in \textit{'Computation of the magnetic field model'} but with the extra free parameter, $r_{0}$. We found that for all models where $r_{0}< 1.5$ kpc, the bisymmetric spiral model does not provide any magnetic field configuration compatible with our observations. For $r_{0} > 1.5$ kpc, the bisymmetric spiral model converges to an axisymmetric spiral morphology within the central $3$ kpc, and both models axisymmetric and bisymmetric are not distinguished.  Because the bisymmetric and axisymmetric spiral field models provide similar solutions for the magnetic field morphology within the central $3$ kpc of Centaurus A, and the axisymmetric spiral field model has less free parameters, we use the axisymmetric spiral model results for our analysis.

The thermal emission of the molecular disk of Centaurus A has been modeled using warped disk configurations \cite{NBT1992,Q1992,QGF1993,Q2006a}. This model consists of tilted concentric rings of material that predict the structure of the warm and cold dust, as well as the velocity fields of the gas in the galaxy disk. Although warped disk models provide compatible solutions for total intensity and spectroscopic observations of several tracers, the concentric and tilted rings do not provide a physical model for magnetic field configurations. Thus, magnetic field configurations based on purely thermal emission or spectroscopic analysis are not considered.

\subsection{Expected emissive polarization from the ISM.} In the following section, we show that our observations trace the magnetic field morphology by means of thermal emission by magnetically aligned dust grains. We can estimate the expected emissive polarization based on previous measurements of the absorptive polarization. At 2.2 \um, the typical degree of polarization in the galaxy disk is P$_{K} \sim 2$\% with a visual extinction of $A_{V} = 7$ mag. ($\tau_{K} = 0.1A_{V} = 0.7$) \cite{capetti2000,packham1996,jones2000}. Using the typical extinction curve (The extinction curve used for the optical depth conversion can be found at https://www.astro.princeton.edu/\~draine/dust/dustmix.html) of the Milky Way for R$_{V} = 3.1$ \cite{WD2001}, we estimate the optical depth at $89$ \um~to be $\tau_{89} = 1.18 \times 10^{-2}  \tau_{K} = 8.26 \times 10^{-3}$, which implies that the dust lane is optically thin at 89 \um. Under the optically thin condition, the emissive polarization can be estimated as $P_{89}^{em} = - P_{89}^{abs}/\tau_{89}$, where the negative sign indicates the change of $90^{\circ}$ from absorptive to emissive polarization \cite{aitken2004}. We scaled the extinction curve, which is representative of the absorptive polarization, at $2$ \um~to be P$_{K} = 2$\% to estimate the expected absorptive polarization at $89$ \um, yield P$_{89}^{abs} = 0.024$\%. Finally, we estimate the expected emissive polarization at 89 \um~to be $P_{89}^{em} \sim 2.9$\%. This result is in good agreement with our median $P=3.5\pm1.9$\% across the whole galaxy disk and it shows that our far-IR polarization measurements at 89 \um~arise from thermal emission of magnetically aligned dust grains. These results also suggest that the ISM in the dust lane of Centaurus A is similar to the ISM of our Galaxy.

\textbf{Temperature and column density maps.} In this section, we show the details to compute the temperature and column density maps. To compute the temperature and hydrogen column density maps (Extended Data Figure 7), we registered and binned to a common $3.2$\arcsec resolution the $70-500$ \um~\textit{Herschel} observations taken with PACS and SPIRE instruments. Then, for every pixel we fit an emissivity modified blackbody function with a constant dust emissivity index $\beta = -2.07$ \cite{parkin2012}. We derived the molecular hydrogen optical depth as $N_{H+H_{2}} = \tau/(k\mu m_{H})$, with the dust opacity $k = 0.1$ cm$^{2}$ g$^{-1}$ at $250$ \um~\cite{H1983}, and the mean molecular weight per hydrogen atom $\mu = 2.8$. Temperature and column density values range from $[20-30]$ K and $log(N _{H+H_{2}} [cm^{-2}]) = [20.6-22.06]$, in agreement with \cite{parkin2012}. Note that our maps cover the central $500$ pc, which was removed from the analysis by \cite{parkin2012}. 

\subsection{Thermal polarization vs. Intensity.} In the following section, we discuss the analysis of the selected subregions across the galactic disk of Centaurus A. In the optically thin regime assuming maximal dust alignment along the LOS and ordered B-field orientation, emissive polarization, $P$, is constant with optical depth, $\tau$ (i.e. $P\propto \tau^{0}$). Thus, some level of variations in the B-field geometry along the LOS will decrease the emissive polarization. For a completely random variation of the B-field orientation along the LOS with a well-defined optical depth scale length $\tau$, then $P\propto\tau^{-0.5}$ \cite{jones1989}, where isotropic random variations of the B-fields do not give rise to polarization. However, observations have found that the slope can be steeper than $-0.5$, e.g. molecular clouds such as OMC-1 \cite{chuss2019}, and external galaxies such as NGC 1068 \cite{LR2018a}. Hydromagnetic simulations have derived that variations of dust grain alignment efficiency and turbulence with column density may explain some of these trends in molecular clouds \cite{king2019}, with a lower-limit of $P\propto \tau^{-1}$.

We have plotted (see Extended Data Figure 8) the debiased polarized flux against the total intensity at $89$ \um~because $P-I$ contains selection effects due to the chosen quality cuts of the degree of polarization and intensity. As the polarized flux is defined as $PI = P \times I$, the equivalence is such as a slope $\alpha$ in $P \propto I^{\alpha}$ become $\beta = \alpha + 1$ in $PI \propto I^{\beta} = I^{\alpha+1}$. The $PI-I$ plot (see Figure 4 and Extended Data Figure 8) shows a complex structure.  In the direction of increasing column density (also $I$), $PI$ increases up to $\log(N_{H+H_{2}} [cm^{-2}]) \sim 21.40$ ($I \sim 750$ MJy sr$^{-1}$), and then decreases with an inflection point at $\log(N_{H+H2} [cm^{-2}]) \sim 21.49$  ($I \sim 1000$ MJy sr$^{-1}$). After, $PI$ increases again up to $\log(N_{H+H_{2}} [cm^{-2}]) \sim 21.72$ ($I \sim 2050$ MJy sr$^{-1}$), and then sharply decreases with an inflection point at $\log(N_{H+H_{2}} [cm^{-2}]) \sim 21.76$ ($I \sim 2700$ MJy sr$^{-1}$). The final trend of $PI$ is an increase with increasing column density up to $\log(N_{H+H_{2}} [cm^{-2}]) \sim 22.06$ ($I \sim 9150$ MJy sr$^{-1}$).  

\textit{Outer disk:}  Using the column density cut of $\log(N_{H+H_{2}}~[cm^{-2}])$ $\le~21.49$, we identify this region as the outer disk of the galaxy (black in Figure 4) with a size of $\sim6$ kpc in diameter. From the $P-I$ plot, the bulk of values shows a trend of $P\propto\tau^{-1}$. If only the outer layers of the disk have perfectly aligned dust grains, then the polarized emission will be diluted by additional unpolarized flux, and a $P\propto\tau^{-1}$ is expected. If high velocity dispersion may be present, then the $P-I$ may show a steep decrease as in the molecular disk. 

\textit{Molecular disk:} Using the range of column densities $ 21.49 \le \log(N_{H+H_{2}}~[cm^{-2}])$ $\le 21.76$, we identify this region as the molecular disk (i.e. parallelogram) of a size of $\sim3$ kpc in diameter (blue in Figure 4).  In the $PI-I$ plot, although the upper values follow the maximum $P=6.5$\%, the $PI$  increases with $N_{H+H_{2}}$ with a steep drop at $\log(N_{H+H_{2}} [cm^{-2}]) \sim 21.76$ and at $T\sim30$ K. We find that the velocity dispersion distribution with a mean of $\sim10$ km s$^{-1}$ is spatially correlated with a trend of $P \propto \tau^{-1}$. The velocity dispersion distribution peaking at $\sigma_{v,^{12}CO(1-0)} \sim20$ km s$^{-1}$ is associated with the steep decrease in the $PI-I$ plot, which creates a bump at the end of the $P-I$ plot.

\textit{Core and low polarized regions:} This region contains several different physical structures. Because we are focused on studying the magnetic fields in the galaxy disk, we only spatially identify the polarized structures in this region. The values with high $PI$, $I$, and $N_{H+H_{2}}$ are identified as the core of Centaurus A at the location of the AGN. The values with intermediate $PI$, and high $I$ and $N_{H+H_{2}}$, are identified as star forming regions in the molecular disk. The values with low $PI$, and high $I$ and $N_{H+H_{2}}$ are identified as low polarized regions mostly located in the central $500$ pc around the AGN. 

\subsection{Thermal polarization vs. $^{12}$CO(1-0) velocity dispersion.} In this section, we describe the data analysis using the $^{12}$CO(1-0) observations. We use the $^{12}$CO(1-0) emission line presented by \cite{E2019} to estimate the velocity dispersions of the galaxy disk of Centaurus A. These observations provide an angular resolution of $2.88 \times 1.67$\arcsec ($46.1 \times 26.9$ pc$^{2}$), which will allow us to estimate the kinematics at the turbulence scales of the galaxy disk. Observations were acquired from the ALMA Archive using the Program ID 2013.1.083.3.S ($^{12}$CO(1-0) ALMA data can be found at http://telbib.eso.org/?bibcode=2019ApJ...887...88E; PI: Espada, D.). We compute the moments using \textsc{immoments} task in \textsc{CASA} with a $1\sigma$ clip, where $\sigma = 68.5$ mJy/beam. Moments were smoothed to the beam size, $7.8$\arcsec, of the HAWC+ observations. The integrated $^{12}$CO(1-0) line (moment 0) and velocity dispersion (moment 2) are shown in Extended Data Figure 7. Using the several physical components distinguished in Figure 4, we measure the median velocity dispersion (see Extended Data Figure 9) across the molecular disk (parallelogram) to be $\sigma_{v,CO}^{MD} = 18.4\pm9.2$ km s$^{-1}$, and outer disk to be $\sigma_{v,CO}^{OD} = 6.4\pm6.0$ km s$^{-1}$ (Extended Data Figure 10). \cite{E2019} estimated a velocity dispersion of $\sim15$ km s$^{-1}$, and $\sim5$ km s$^{-1}$ in the molecular disk and outer disk, respectively.  These two populations of velocity dispersion may be biasing the mean $\sigma_{v,^{12}CO(1-0)} = 18.4\pm9.2$ km s$^{-1}$ to larger values than those typically found, $\sim 8$ km s$^{-1}$, in nearby galaxies.

\subsection{Power-law fits.}   We have fit a power-law, $y \propto x^{\alpha}$, for each of the plots and physical regions shown in Figure 5. Extended Data Figure 9 shows the power-law indexes, $\alpha$, of the fits for each physical region.

\subsection{Alternative scenario.} Another possibility for the observed B-field could arise from field compression and amplification due to density-shocks, where density-shocks are discontinuities in the flow due to change of gas properties (density, temperature, velocity, etc.). These density-wave shocks bend the B-fields along the shocks, reducing the angular dispersion, which would make the general B-field morphology indistinguishable from regular large-scale fields \cite{DP2008}. Therefore, small-scale anisotropic B-fields arising from compression and amplification due to density-shocks are inconsistent with our measured angular dispersions.

\end{methods}

\begin{addendum}
 \item[Data Availability] The data that support the plots within this paper and other findings of this study are available from https://galmagfields.com or from the corresponding author upon reasonable.
 \item[Code Availability] The code that support the algorithms within this paper and other findings of this study are available from https://github.com/galmagfields or from the corresponding author upon reasonable.
 \end{addendum}

\begin{addendum}
\item[Extended Data Figure 1:] \textbf{Summary of OTFMAP polarimetric observations.}
\item[Extended Data Figure 2:] \textbf{Polarization measurements of the several regions of the galactic disk. }
\item[Extended Data Figure 3:] \textbf{Physical regions based on B-field orientation and degree of polarization.} Histograms of P (\textbf{a}) and PA (\textbf{b}) of polarization for measurements with $P/\sigma_{P} \ge 3.0$. Three distinct regions are found for the PA of polarization, which are identified with the west (orange), east (red) and low polarized (black) regions. The boundaries of each region are shown with vertical black dashed lines. \textbf{c}, The spatial correspondence of the three regions identified using the  PA distributions are shown with the same colors as the plots at \textbf{b}. The total intensity contours are shown as in Figure 1. A legend polarization of $10$\% (black) and beam size of $7.8$\arcsec (red circle) are shown. 
\item[Extended Data Figure 4:] \textbf{Magnetic field of the central $50\arcsec \times 50\arcsec$ ($0.8 \times 0.8$ kpc$^{2}$) of Centaurus A.} \textbf{a}, Total flux (colorscale) with overlaid B-field orientations (white lines). \textbf{b}, Polarized flux (colorscale) with overlaid B-field orientation (white lines). A legend polarization of $5$\% (black) and beam size of $7.8$\arcsec (red circle) are shown. 
\item[Extended Data Figure 5:] \textbf{Parameters of the magnetic field morphological model.}
\item[Extended Data Figure 6:] \textbf{Posterior distributions of the magnetic field morphological model.} A reference of the parameter definitions, used priors, and median values is shown in Extended Data Figure 5.
\item[Extended Data Figure 7:] \textbf{Polarization map vs physical parameters.} Temperature (\textbf{a}) and column density (\textbf{b}) maps of Centaurus A with overlaid B-field orientation (while lines) with $P/\sigma_{p} \ge 2.5$ and $PI/\sigma_{PI} \ge 2$. Temperature contours start at $20$ K increasing in steps of $0.5$ K, and column density density contours start at $\log(N_{H+H2} [cm^{-2}]) = 20.6$ increasing in steps of 0.1. $^{12}CO(1-0)$ integrated line emission (\textbf{c}) and velocity dispersion (\textbf{d}) of the warped disk of Centaurus A with overlaid B-field orientation (white lines) with $P/\sigma_{p} \ge 2.5$ and $PI/\sigma_{PI} \ge 2$.
\item[Extended Data Figure 8:] \textbf{Polarized flux vs. total intensity plots.} $P-I$ and $PI-I$ plots at 89 \um~vs temperature (\textbf{a}, \textbf{b}) and column density (\textbf{c}, \textbf{d}). The  trend of the bulk of the $P-I$ plot, $P \propto \tau^{-1}$, is shown as a black solid line in panels \textbf{a} and \textbf{c}. The uncertainties of the debiased polarized intensity in plots \textbf{b} and \textbf{d} are shown. The blue dotted vertical lines at $I = 1000$ and $2700$ MJy sr$^{-1}$ show the limits of the three physical regions found in this analysis. The black dotted lines in panels \textbf{b} and \textbf{d} show the maximum expected polarization, $P\propto I^{0}$ = 15, 6.5, and 1.5\% for each of these physical regions, respectively.
\item[Extended Data Figure 9:] \textbf{Power-law index of plots from Figure 5.}
\item[Extended Data Figure 10:] \textbf{Velocity dispersion of the outer and molecular disk.} $^{12}CO(1-2)$ velocity dispersion histograms of the outer disk (red) and molecular disk (blue) as identified in Figure 4. The median (solid line) and $1\sigma$ (dashed line) are shown for each physical structure. These values correspond to $\sigma_{v,^{12}CO(1-0)} = 18.4\pm9.2$ (km s$^{-1}$), and  $\sigma_{v,^{12}CO(1-0)} = 6.4\pm6.0$ (km s$^{-1}$) for the molecular disk and outer disk, respectively.
\end{addendum}


\begin{figure*}[ht!]
\includegraphics[angle=0,scale=0.4]{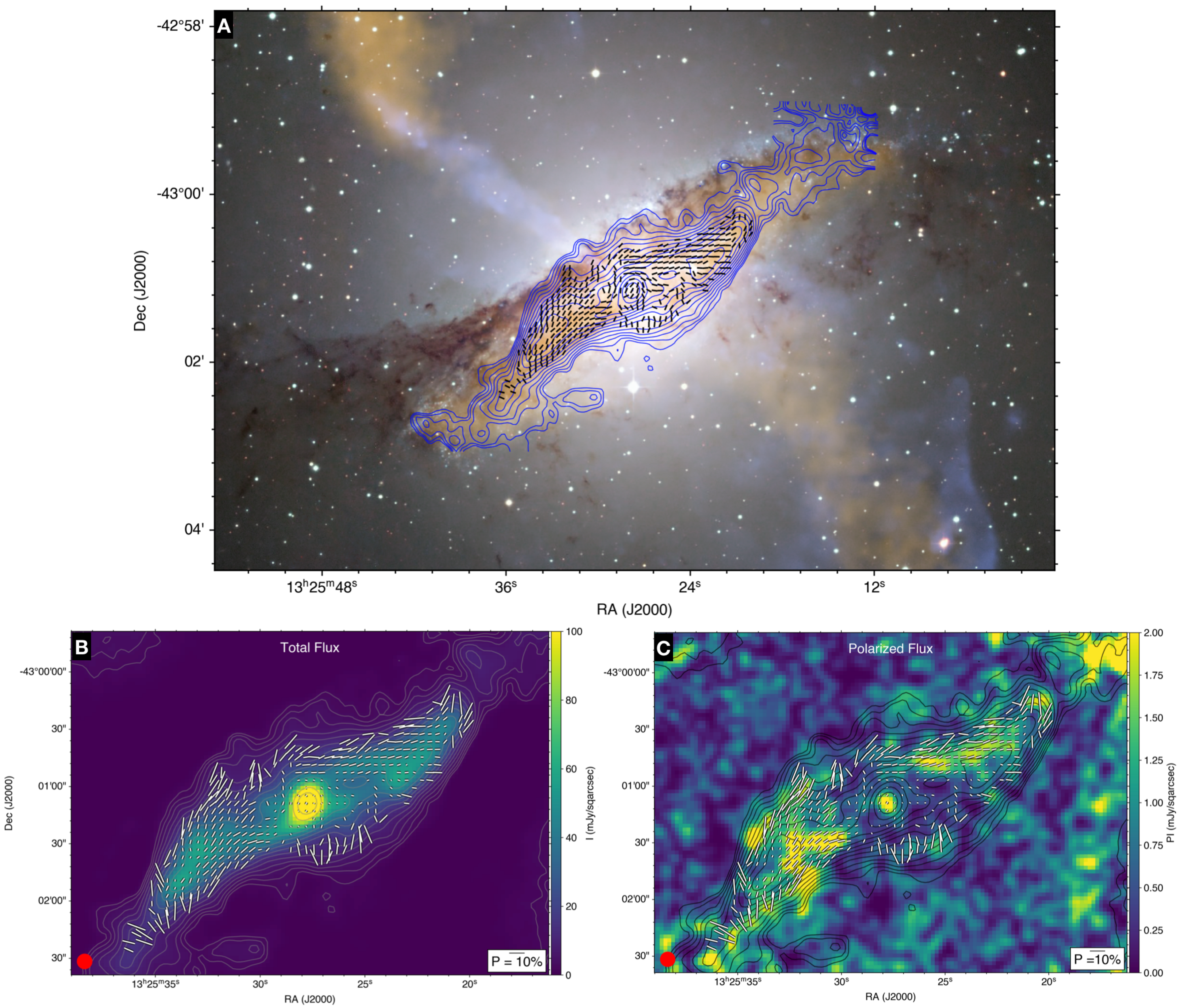}
\caption{
\label{fig:fig1}}
\end{figure*}

\begin{figure*}[ht!]
\includegraphics[angle=0,scale=0.4]{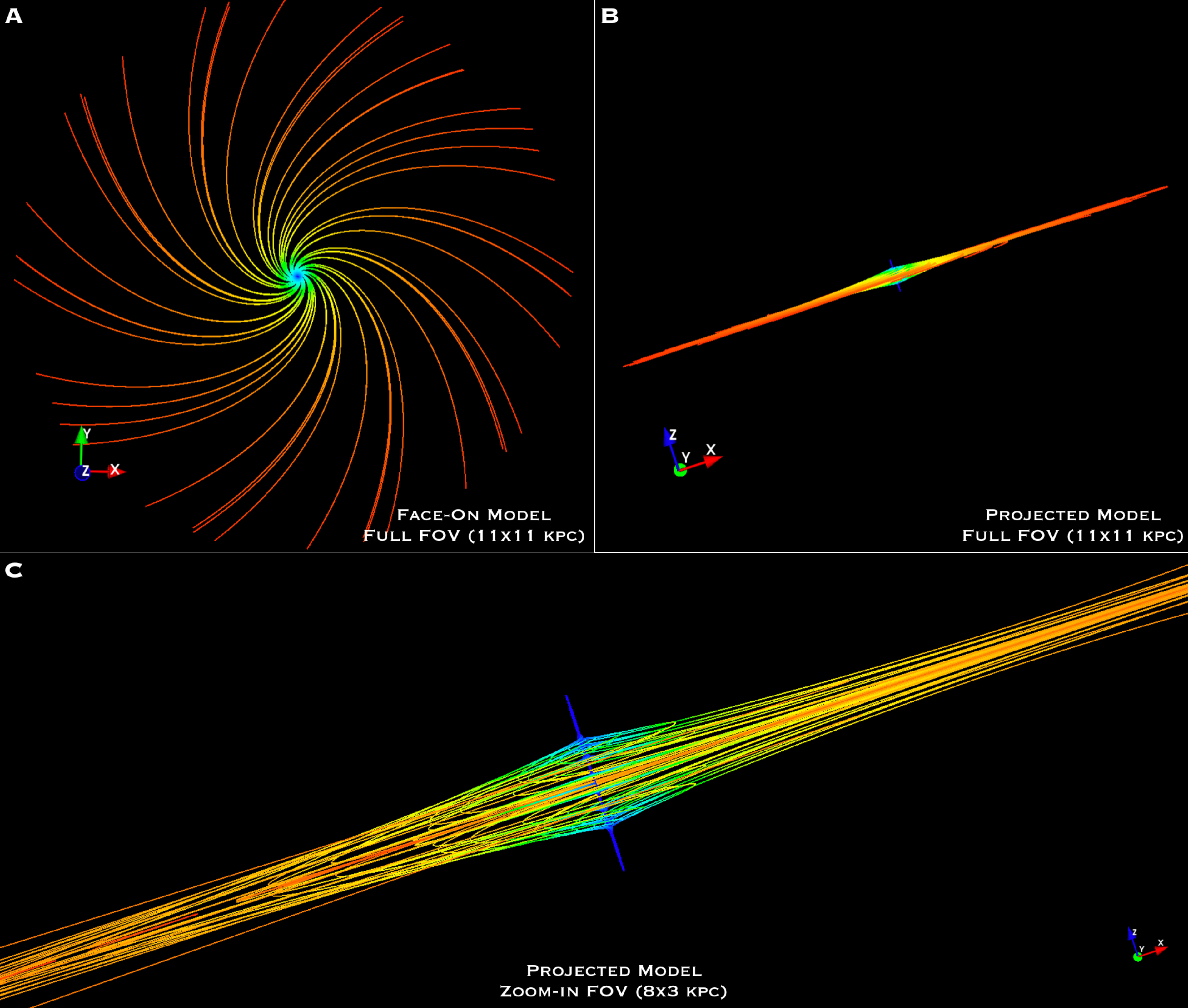}
\caption{
\label{fig:fig2}}
\end{figure*}

\begin{figure*}[ht!]
\includegraphics[angle=0,scale=0.45]{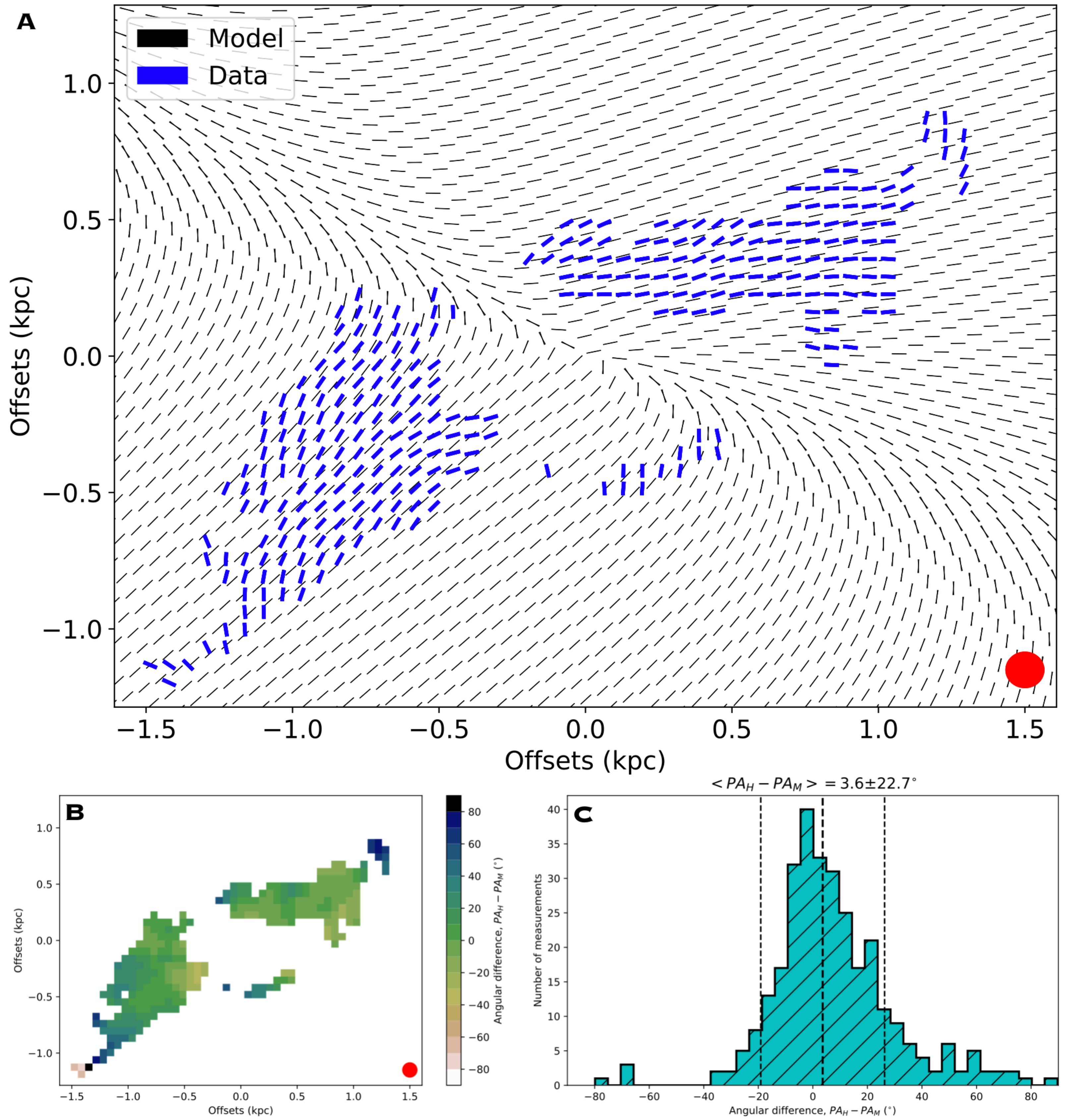}
\caption{
\label{fig:fig3}}
\end{figure*}

\begin{figure*}[ht!]
\includegraphics[angle=0,scale=0.4]{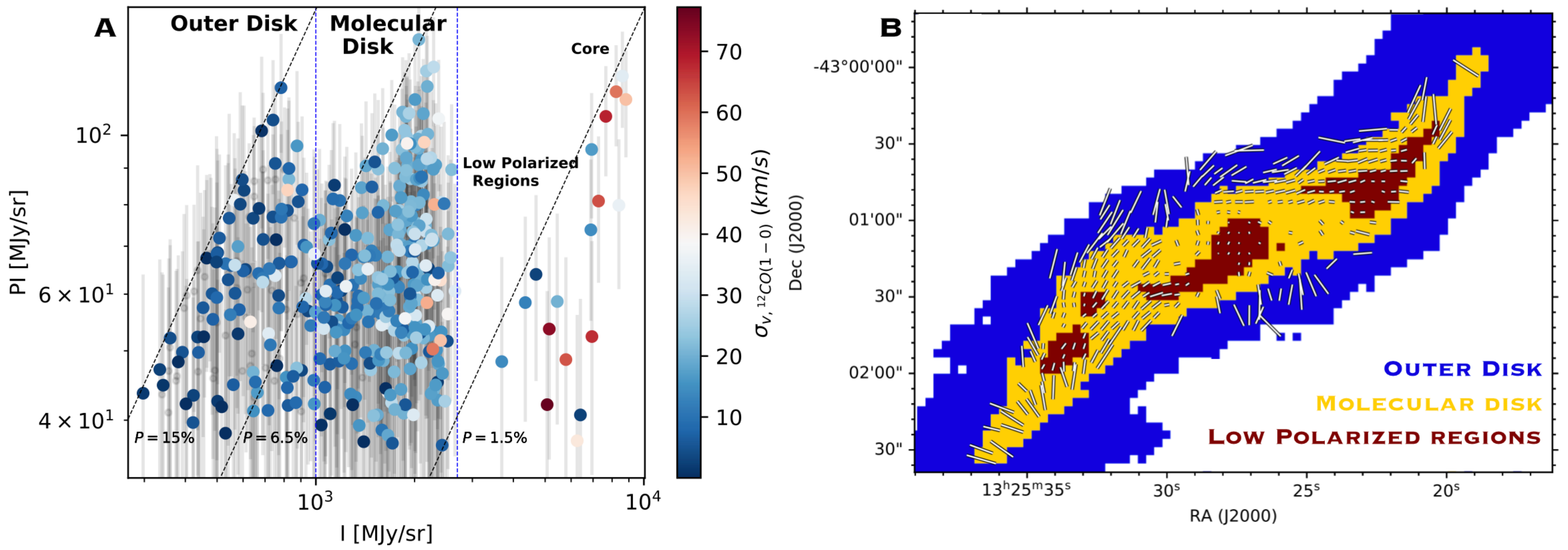}
\caption{
\label{fig:fig4}}
\end{figure*}

\begin{figure*}[ht!]
\includegraphics[angle=0,scale=0.55]{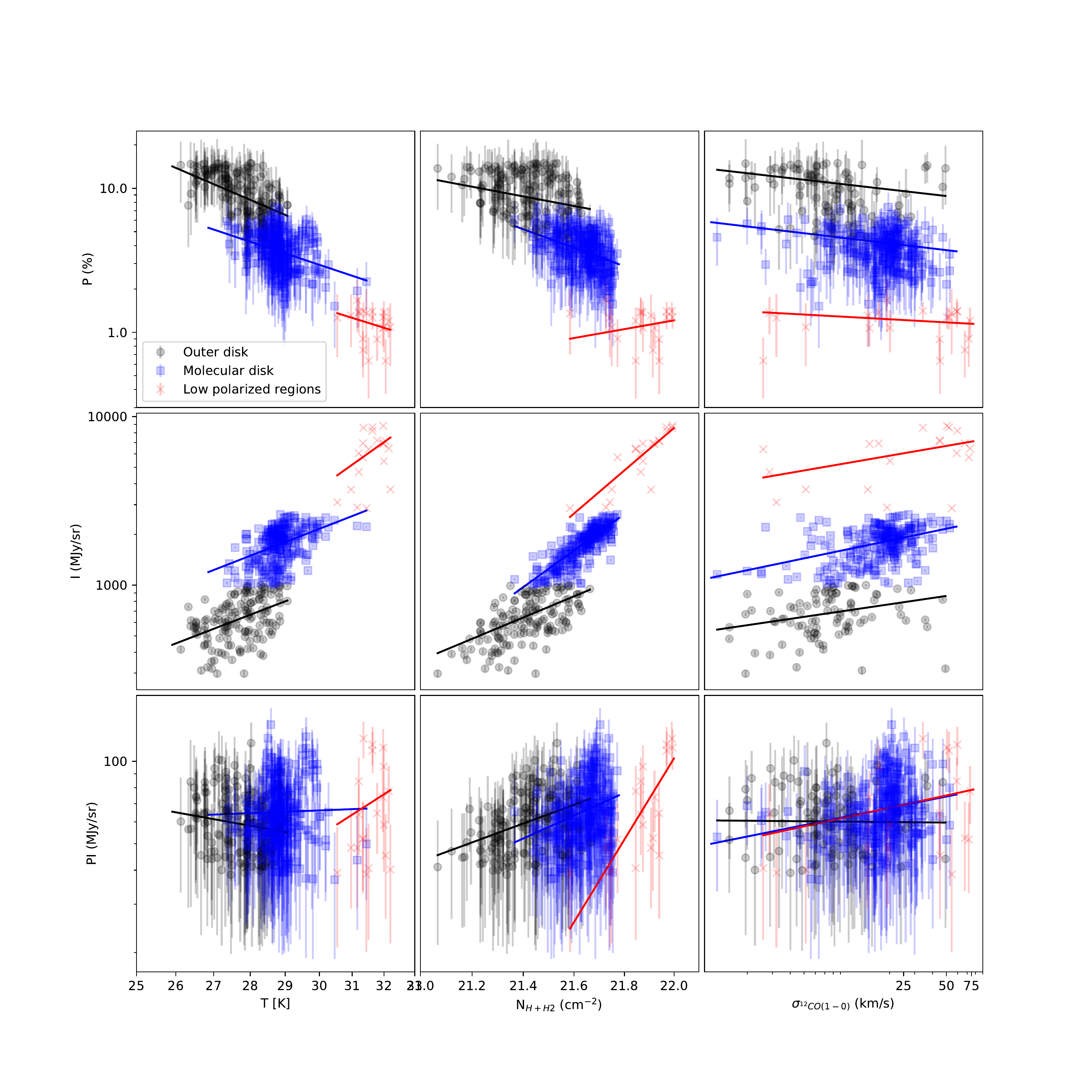}
\caption{ \label{fig:fig5}}
\end{figure*}

\begin{figure*}[ht!]
\includegraphics[angle=0,scale=0.6]{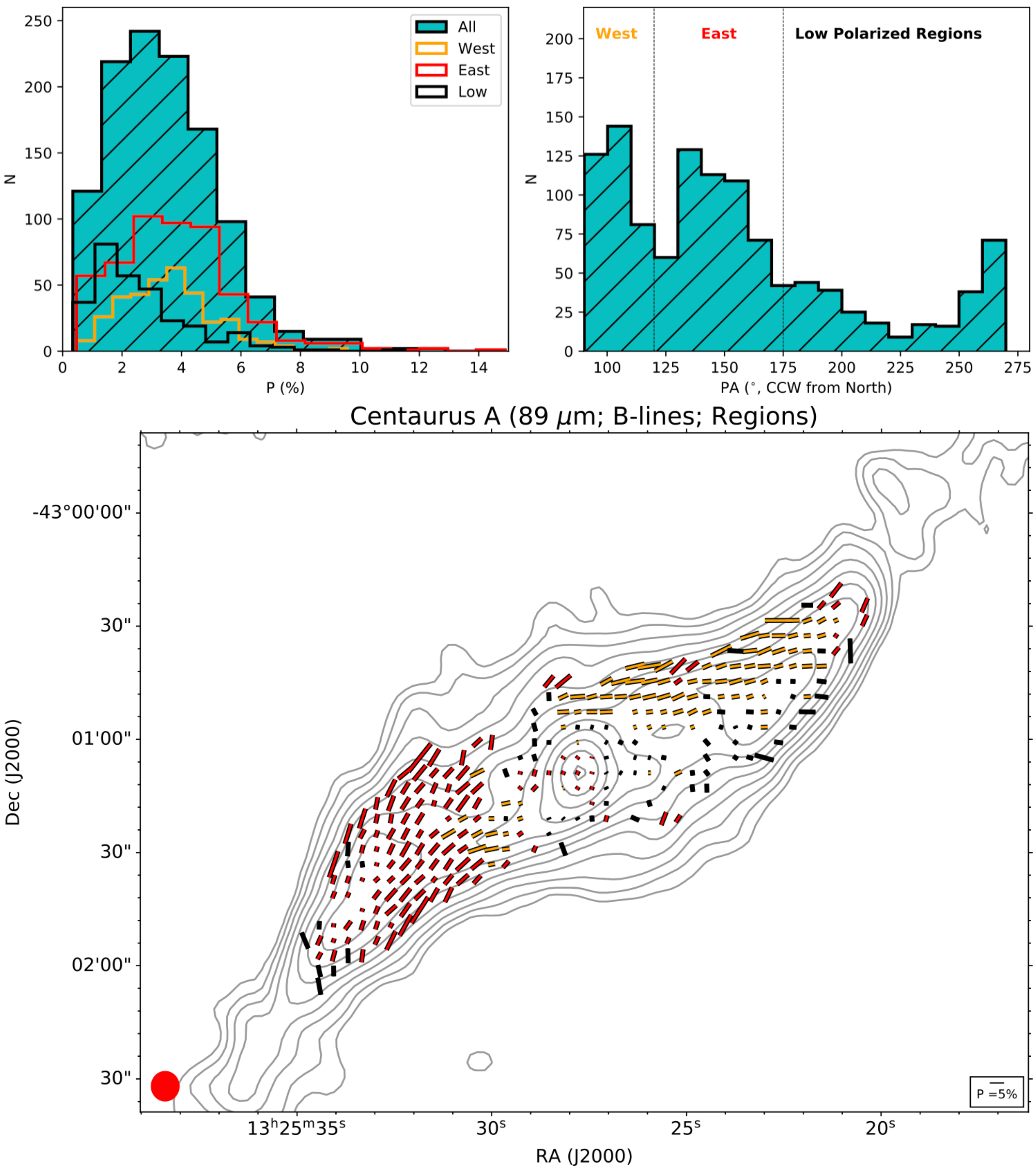}
\caption{Extended Data Figure 3
\label{fig:fig6}}
\end{figure*}

\begin{figure*}[ht!]
\includegraphics[angle=0,scale=0.4]{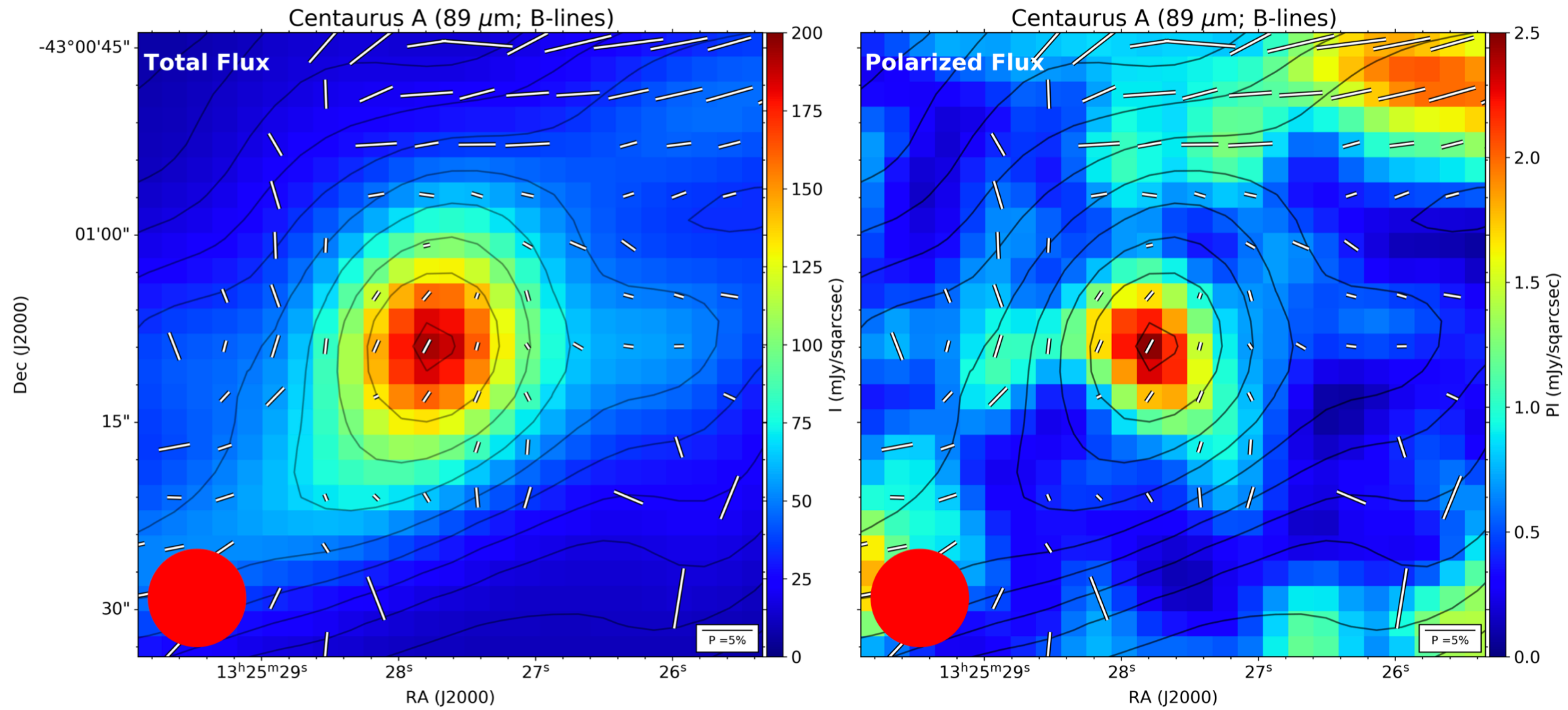}
\caption{Extended Data Figure 4.   
\label{fig:fig7}}
\end{figure*}

\begin{figure*}[ht!]
\includegraphics[angle=0,scale=0.55]{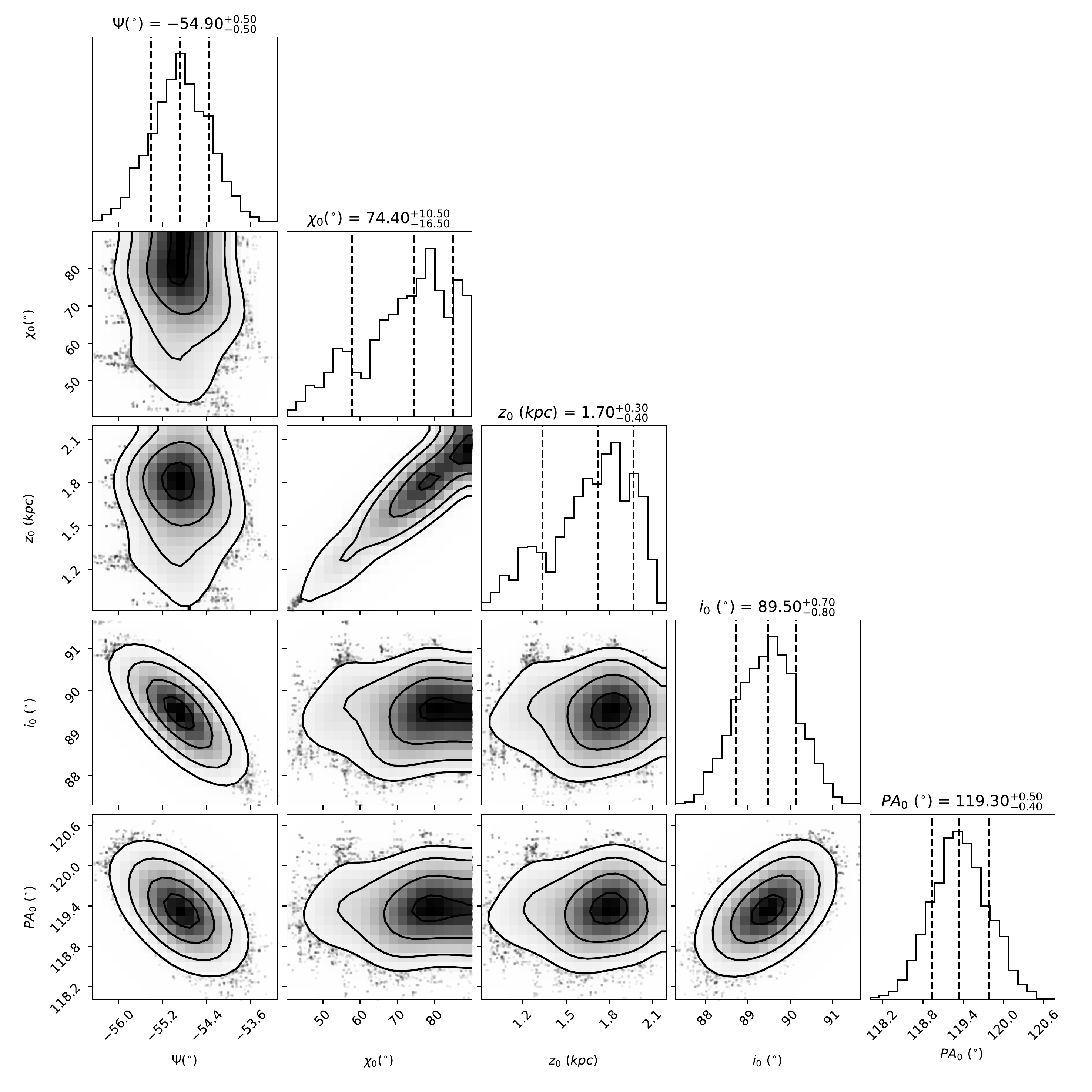}
\caption{Extended Data Figure 6. 
\label{fig:fig8}}
\end{figure*}

\begin{figure*}[ht!]
\includegraphics[angle=0,scale=0.4]{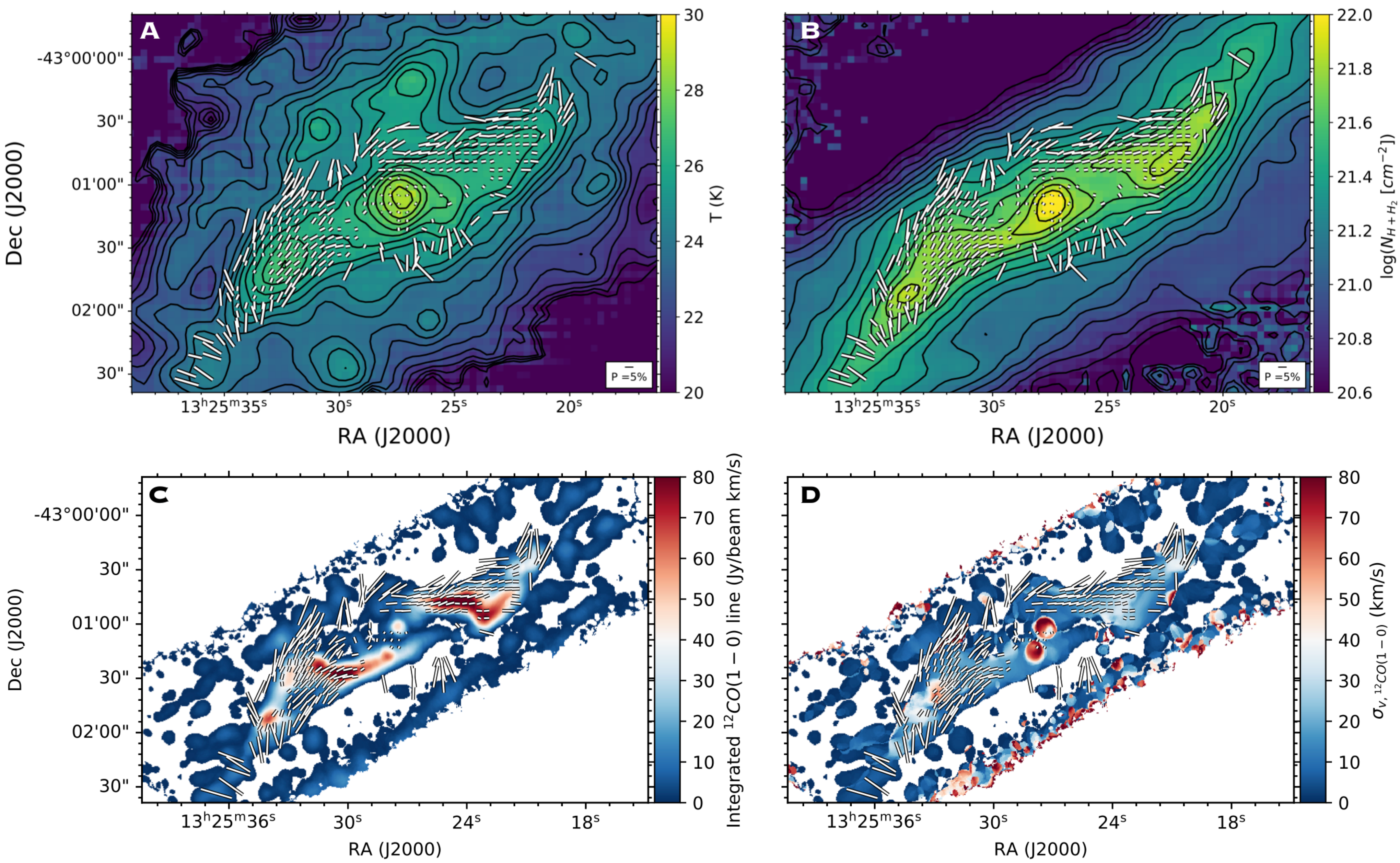}
\caption{Extended Data Figure 7. 
\label{fig:fig9}}
\end{figure*}

\begin{figure*}[ht!]
\includegraphics[angle=0,scale=0.4]{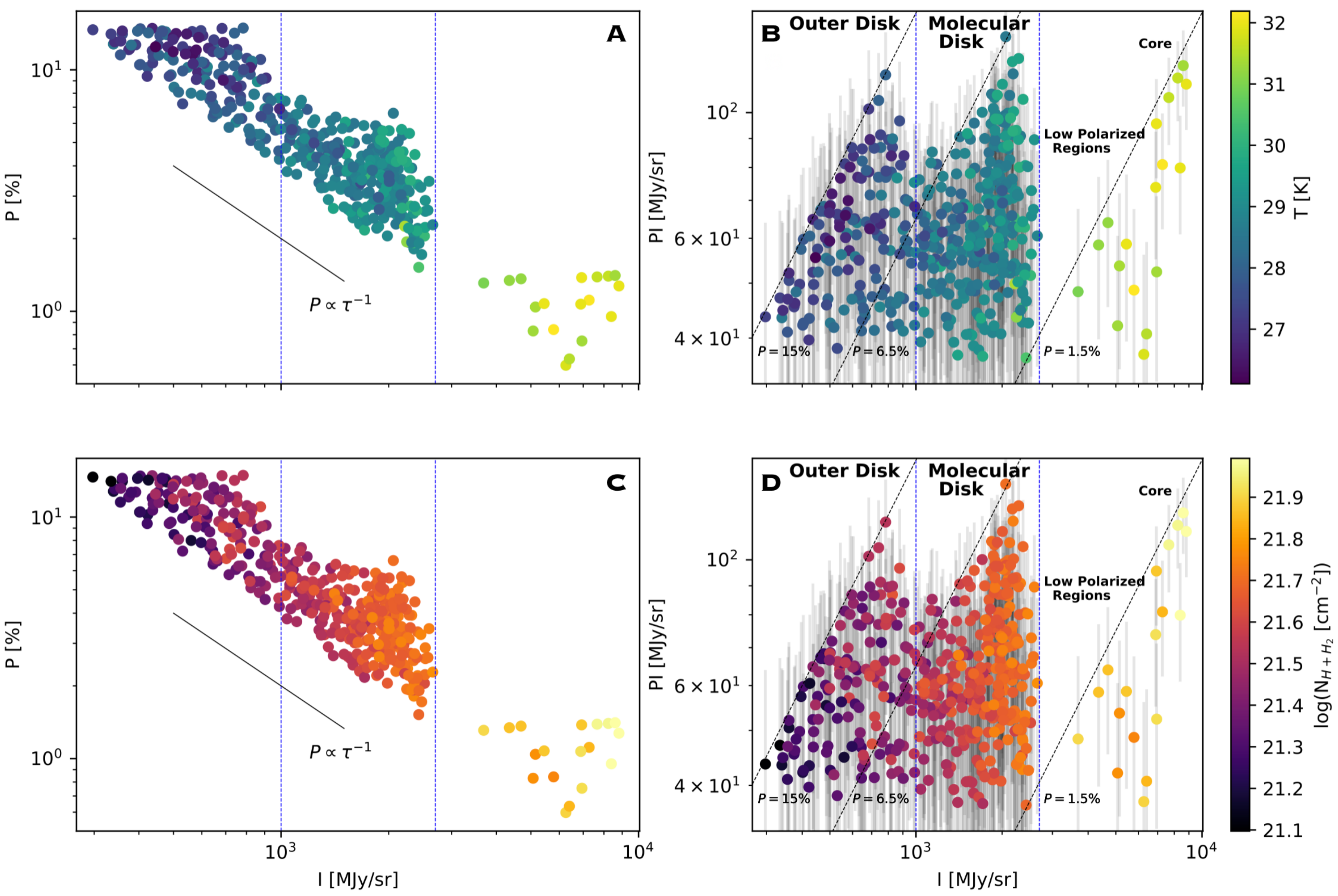}
\caption{Extended Data Figure 8.
\label{fig:fig10}}
\end{figure*}

\begin{figure}[ht!]
\includegraphics[angle=0,scale=0.8]{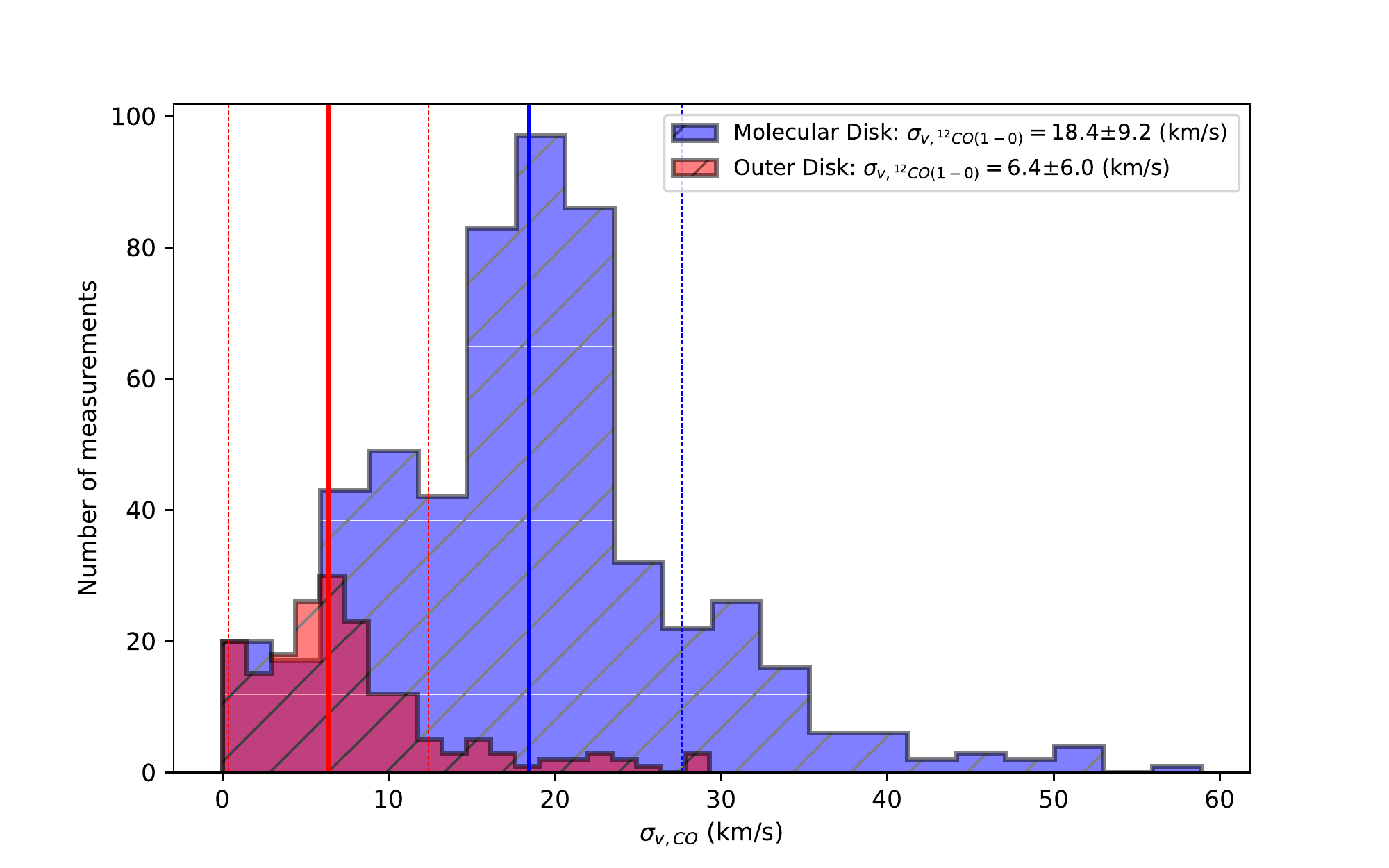}
\caption{Extended Data Figure 10.
\label{fig:fig11}}
\end{figure}


\begin{table}
\caption{Medians of the physical parameters of each region identified in Figure 4}
\centering
\begin{tabular}{lcccc}
\hline\hline
Parameter & Outer Disk  &   Molecular Disk &   Low Polarized \\
\hline
T (K)							&	$27.8\pm0.7$	&	$28.8\pm0.8$	&	$31.4\pm0.9$ \\
$\sigma_{v,^{12}CO(1-0)}$ (km/s)	&	$6.4\pm6.0$	&	$18.4\pm9.2$	&	$34\pm4$ \\
P (\%)						&	$9.5\pm3.3$	&	$3.9\pm2.4$	&	$1.2\pm0.6$ \\
PI (MJy/sr)					&	$55\pm13$	&	$63\pm18$	&	$71\pm26$ \\
I (MJy/sr)						&	$627\pm264$	&	$1761\pm554$	&	$6500\pm1876$ \\
\hline
\end{tabular}
\label{tab:table1}
\end{table}

\begin{table}
\caption{Summary of OTFMAP polarimetric observations.}
\centering
\begin{tabular}{cccccccccc}
\hline\hline
Wavelength &	Bandwidth	& Beam Size	 & Scan Rate & Scan Phase & Scan Amplitude  \\ 
(\um) & (\um) & (\arcsec) & (\arcsec/s) & ($^{\circ}$) & (EL $\times$ XEL; \arcsec)  \\
\hline
89	&	17.0	&	7.80	&	100	&	0	&	180 $\times$ 120	\\
\hline\hline	
Scan Duration & \#Sets & t$_{on-source}$ \\
(s)  & & (s)\\
\hline
100	&	8	& 3200 \\
\hline
\end{tabular}
\label{tab:table2}
\end{table}


\begin{table}
\caption{Polarization measurements of the several regions of the galactic disk. $\star\sigma_{P}$ and $\sigma_{PA}$ correspond to the dispersion of the  measurements of the degree and PA of polarization per region.}
\centering
\begin{tabular}{cccccccccc}
\hline\hline
Regions & PA$_{B}$  &  $\sigma_{PA_{B}}\star$ &  P & $\sigma_P\star$  \\
				& ($^{\circ}$)  &  ($^{\circ}$)   & (\%)   & (\%)  \\ 
\hline\hline
West		&	$105$	&	$8.6$	&	$3.5$	&	$1.6$	\\
East		&	$147$	&	$15.5$	&	$3.6$	&	$2.0$	\\
Low-P	&	$-$		&	$28.9$	&	$1.5$	&	$1.7$	\\
All		&	$-$		&	$-$		&	$3.5$	&	$1.9$	\\
\hline 
\end{tabular}
\label{tab:table3}
\end{table}

\begin{table}
\caption{Parameters of the magnetic field morphological model}
\centering
\begin{tabular}{cccccccccc}
\hline\hline
Parameter & Symbol  &   Priors &   Median Posteriors \\
\hline
Pitch angle ($^{\circ}$)		& $\Psi_{0}$	&	$[-90,0]$		&	$-54.9^{+0.5}_{-0.5}$	\\
Radial pitch angle ($^{\circ}$) 	& $\chi_{0}$	&	$[0,90]$		&	$74.4^{+10.5}_{-16.5}$	\\
Vertical scale (kpc)			& $z_{0}$		&	$[0,10]$ 		&	$1.7^{+0.3}_{-0.4}$ 		\\
Inclination ($^{\circ}$)		& $i$			&	$[0,90]$		&	$89.5^{+0.7}_{-0.8}$		\\
Tilt angle ($^{\circ}$)			& $\theta$		&	$[0,180]$		&	$119.3^{+0.5}_{-0.4}$		\\
\hline 
\end{tabular}
\label{tab:table4}
\end{table}

\begin{table}
\caption{Power-law index of plots from Fig. \ref{fig:fig5}}
\centering
\begin{tabular}{cccccccccc}
\hline\hline
Parameters & Regions  & T  &  $N_{H}$ &   $\sigma_{v,^{12}CO(1-0)}$ \\
\hline
$P$ 	&	Outer disk		&	$-6.89\pm0.47$		&	$-16.49\pm1.24$	&	$-0.18\pm0.04$\\
 	&	Molecular disk	&	$-5.35\pm0.49$		&	$-31.91\pm1.21$	&	$-0.13\pm0.03$\\
	&	Low Polarized	&	$-5.00\pm10.12$	&	$15.39\pm12.06$	&	$-0.02\pm0.05$\\
\hline
$I$	&	Outer disk		&	$5.27\pm0.47$		&	$31.07\pm5.99$	&	$0.12\pm0.02$\\
 	&	Molecular disk	&	$5.33\pm0.17$		&	$53.87\pm2.78$	&	$0.18\pm0.01$\\
	&	Low Polarized	&	$9.73\pm10.33$	&	$63.79\pm52.19$	&	$0.21\pm0.01$\\
\hline
$PI$	&	Outer disk		&	$-1.52\pm0.36$		&	$16.94\pm5.77$	&	$-0.01\pm0.04$\\
 	&	Molecular disk	&	$0.33\pm0.36$		&	$20.71\pm9.79$	&	$0.10\pm0.03$\\
	&	Low Polarized	&	$5.41\pm21.32$	&	$75.15\pm69.91$	&	$0.11\pm0.05$\\
\hline
\end{tabular}
\label{tab:table5}
\end{table}

\end{document}